\documentclass[prb,twocolumn,superscriptaddress]{revtex4-2}

\usepackage{graphicx,color}
\usepackage{amsmath}
\usepackage{amssymb}
\usepackage{natbib}
\usepackage{tabularx}
\usepackage{multirow}
\usepackage[colorlinks=true,linkcolor=blue,citecolor=blue,urlcolor=black]{hyperref}
\usepackage{ulem}

\newcommand{\ndzr}{Nd$_2$Zr$_2$O$_7$}
\newcommand{\ndhf}{Nd$_2$Hf$_2$O$_7$}
\newcommand{\ndzrx}{Nd$_2$(Zr$_{1-x}$Ti$_x$)$_2$O$_7$}

\newcommand{\ybti}{Yb$_2$Ti$_2$O$_7$}
\newcommand{\erti}{Er$_2$Ti$_2$O$_7$}

\newcommand{\tbti}{Tb$_2$Ti$_2$O$_7$}

\newcommand{\nd}{Nd$^{3+}$}
\newcommand{\TN}{$T_{\rm N}$}

\newcolumntype{P}{>{\centering\arraybackslash}p}
\newcolumntype{Y}{>{\centering\arraybackslash}X}

\begin{document}

\title{Field-temperature phase diagram of the enigmatic \ndzrx\ pyrochlore magnets}

\author{M. L\'eger}
\email[]{melanie.leger@neel.cnrs.fr}
\affiliation{Institut N\'eel, CNRS and Universit\'e Grenoble Alpes, 38000 Grenoble, France}
\affiliation{Laboratoire L\'eon Brillouin, Universit\'e Paris-Saclay, CNRS, CEA, CE-Saclay, F-91191 Gif-sur-Yvette, France}
\author{E. Lhotel}
\email[]{elsa.lhotel@neel.cnrs.fr}
\affiliation{Institut N\'eel, CNRS and Universit\'e Grenoble Alpes, 38000 Grenoble, France}
\author{E. Ressouche}
\affiliation{IRIG, CEA and Universit\'e Grenoble Alpes, CEA Grenoble, F-38054 Grenoble, France}
\author{K. Beauvois}
\affiliation{IRIG, CEA and Universit\'e Grenoble Alpes, CEA Grenoble, F-38054 Grenoble, France}
\affiliation{Institut Laue Langevin, F-38042 Grenoble, France}
\author{F. Damay}
\affiliation{Laboratoire L\'eon Brillouin, Universit\'e Paris-Saclay, CNRS, CEA, CE-Saclay, F-91191 Gif-sur-Yvette, France}
\author{C. Paulsen}
\affiliation{Institut N\'eel, CNRS and Universit\'e Grenoble Alpes, 38000 Grenoble, France}
\author{A. Al-Mawla}
\affiliation{Institut N\'eel, CNRS and Universit\'e Grenoble Alpes, 38000 Grenoble, France}
\author{E. Suard}
\affiliation{Institut Laue Langevin, F-38042 Grenoble, France}
\author{M. Ciomaga Hatnean}
\affiliation{Department of Physics, University of Warwick, Coventry, CV4 7AL, United Kingdom}
\author{G. Balakrishnan}
\affiliation{Department of Physics, University of Warwick, Coventry, CV4 7AL, United Kingdom}
\author{S. Petit}
\email[]{sylvain.petit@cea.fr}
\affiliation{Laboratoire L\'eon Brillouin, Universit\'e Paris-Saclay, CNRS, CEA, CE-Saclay, F-91191 Gif-sur-Yvette, France}

\begin{abstract}
By combining neutron scattering and magnetization measurements down to 80 mK, we determine the $(H,T)$ phase diagram of the \ndzrx\ pyrochlore magnet compounds. In those samples, Zr is partially substituted by Ti, hence tuning the exchange parameters and testing the robustness of the various phases. In all samples, the ground state remains ``all in / all out'', while the field induces phase transitions towards new states characterized by ``2 in -- 2 out'' or ``1 out -- 3 in / 1 in -- 3 out'' configurations. These transitions manifest as metamagnetic singularities in the magnetization vs field measurements. Strikingly, it is found that moderate substitution reinforces the stability of the ``all in / all out'' phase: the N\'eel temperature, the metamagnetic fields along with the ordered magnetic moment are higher in substituted samples with $x <$ 10\%.
\end{abstract}

\maketitle

%%%%%%%%%%%%%%%%%%%%%%%%%%%%
%INTRODUCTION 
%%%%%%%%%%%%%%%%%%%%%%%%%%%%

\section{Introduction}

The last decades of research in the field of condensed matter have seen the emergence of a rich and new physics, going beyond the N\'eel paradigm and transcending conventional descriptions based on Landau's theory. Frustrated magnetism has largely contributed to these developments, pointing to the existence of new states of matter, such as spin liquids or Coulomb phases. The interest in these states originates from the fact that they host fractional excitations, are described by emergent gauge fields and exhibit large-scale quantum entanglement \cite{Lacroix11,Gingras14,Gardner10}. 

The pyrochlore network, built from tetrahedra joined by their vertices has played a key role in this field. In particular, Ising spins located at the summits of these tetrahedra and coupled by ferromagnetic interactions form an unconventional state called spin ice \cite{Harris97}. The spins remain disordered, but nevertheless obey a local rule known as the ``2 in -- 2 out'' rule, which stipulates that each tetrahedron must have two incoming and two outgoing spins, in close analogy with the disorder of hydrogen atoms in water ice. It results in a residual entropy at zero temperature \cite{Ramirez99}, as well as in a peculiar local organization of the spins. The latter can be observed by neutron scattering, and manifests as an elastic pattern in reciprocal space with bow tie singularities also called pinch points \cite{Henley05,Isakov04,Fennell2009}. Furthermore, the ice rule can be interpreted as the zero divergence condition of an emergent magnetic field (${\bf \nabla} \cdot {\bf B} = 0$), a mapping that has allowed considerable theoretical development.

In most of the pyrochlore materials of interest today, the spin is due to the 4f magnetic moment of a rare-earth. These compounds have been abundantly investigated in the literature \cite{Gardner10}, with the description of a large variety of ground states arising from the interplay between magnetic exchange couplings, dipolar interactions, and single-ion anisotropy of the magnetic moment. It is worth mentioning spin ices, but also spin liquids, fragmented states and apparently more conventional magnets.
%It is worth mentioning spin ices, as \hoti\ or \dyti\ \cite{Harris97, Ramirez99}, but also spin liquids (\cesn\ \cite{Sibille20}, \prhf\ \cite{Sibille18}, \przr\ \cite{Kimura13,PetitPrZr16,Martin17}, \tbti\ \cite{Gardner99}), fragmented states (\hoir\ \cite{Lefrancois_2017} and \dyir\ \cite{Cathelin20}) and apparently more conventional magnets as \erti\ \cite{Champion03} and \ersn\ \cite{Petit17}.

Among them, \ndzr\ has attracted much attention. As inferred from a positive Curie-Weiss temperature, indicating ferromagnetic interactions, and combined with an Ising anisotropy, a spin ice ground state is expected in this material \cite{Lhotel_2015,Xu_2015}. Indeed, the expected pinch point pattern is observed below $1$ K \cite{Petit_2016}. Upon further cooling, however, the spin ice state gives way to an ``all in / all out'' (AIAO) ground state, with all the spins pointing out or into each tetrahedron, hence highlighting dominant antiferromagnetic interactions \cite{Lhotel_2015, Xu_2015, Benton_2016}. 

Motivated by this very rich physics and by the competition between spin ice and AIAO state \cite{Xu_2020, Leger20}, we investigate in this paper the field - temperature $(H,T)$ phase diagram of this material. Moreover, as documented in many pyrochlore compounds such as \tbti\ \cite{Taniguchi_2013}, \erti\ \cite{Shirai_2017} or \ybti\ \cite{Arpino_2017}, weak substitution, often less than 5\%, can change the physical properties leading to a quick collapse of the ground state. With this in view, the study was extended to the \ndzrx\ family, using two nominal compositions $x=0.025$ and $x=0.1$. These investigations yield a systematic and comprehensive survey of the $(H,T)$ phase diagram for magnetic fields applied along the three main high symmetry directions ($[001]$, $[110]$ and $[111]$) of the pyrochlore lattice.

The paper is organized as follows. In section II, we introduce the main properties of \ndzr\ and the context of the study. In section III and IV, we present the experimental methods and determination of the $(H,T)$ phase diagrams for the three studied samples and the three field directions. Finally, by comparing these results with mean-field calculations, we discuss in section V the involved field induced processes, including the role of domains and metastable states, and address the impact of Ti substitution on the phase diagram.

%%%%%%%%%%%%%%%%%%%%%%%%%%%%%%%%%%%%%%%%
\section{Experimental background and interpretations : a short review}

\begin{figure}[h!]
\includegraphics[width=6cm]{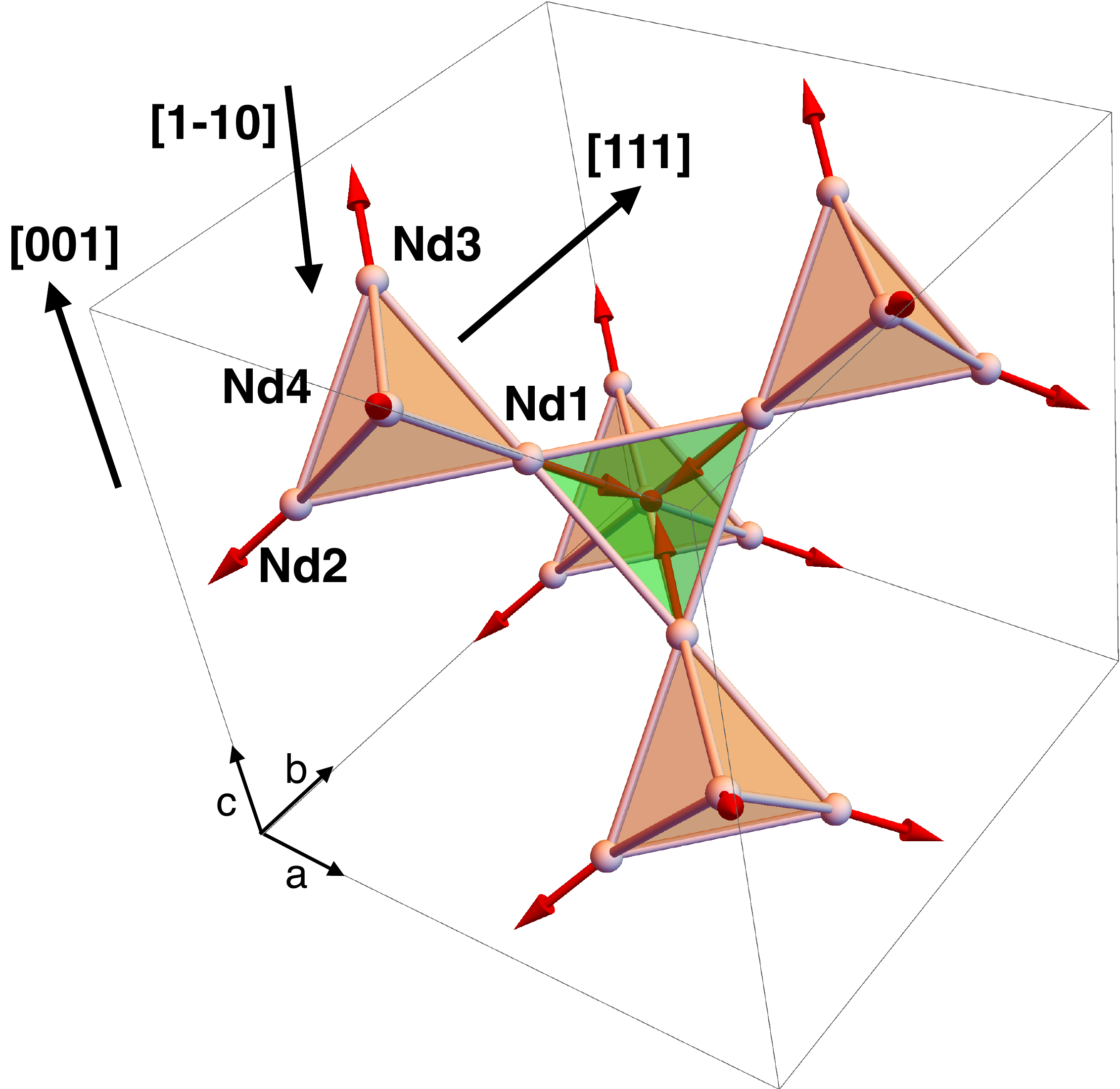}
\caption{\label{Image1} 
Sketch of one configuration of the AIAO ordered state of the Nd pyrochlore lattice in the unit cell of \ndzr. By convention, outer tetrahedra are red, while the central tetrahedron is green. Note that to avoid any confusion between the so-called ``all in / all out" magnetic structure, and the possible magnetic domains associated to this state, we use different notations for the state (AIAO) and for the domains (${\rm \overline{AIAO}}$ or ${\rm \overline{AOAI}}$). The depicted configuration, with red ``all out'' tetrahedra and green ``all in'' tetrahedron, is labeled ${\rm \overline{AOAI}}$. The other configuration, ${\rm \overline{AIAO}}$, would have red ``all in'' tetrahedra and green ``all out'' tetrahedron. The figure also features the \nd\ ions labeled according to Table~\ref{table_ND}. 
}
\end{figure}

Using a combination of magnetization, elastic and inelastic neutron scattering measurements \cite{Lhotel_2015,Petit_2016}, previous investigations of \ndzr\ have shown that the \nd\ magnetic moments exhibit a strong Ising anisotropy along local $\langle$111$\rangle$ axes, together with a dipolar-octupolar nature \cite{Huang14}, different from the standard Kramers doublets studied so far. Furthermore, neutron diffraction has shown that the AIAO antiferromagnetic state (Figure \ref{Image1}) is characterized by a strongly reduced ordered moment \cite{Lhotel_2015,Xu_2015,Opherden_2017}, and coexists with a pinch point pattern typical of spin ice. Nevertheless, inelastic neutron scattering clearly pointed out that, at low temperature, this pattern is not an elastic feature as in classical spin ice, yet is dynamic, shifted to finite energy by an energy $E_o \approx 70~\mu$eV, and that was hence called a dynamical spin ice mode \cite{Petit_2016}. In other terms, the peculiar AIAO order of \ndzr\ appears to be protected by a ``gap" $E_o$ from spin ice. In addition, the spin excitation spectrum revealed by inelastic neutron scattering also encompasses well resolved dispersing features akin to spin wave branches. These experimental results suggested that quantum effects are at play. 

Concomitantly, theoretical studies highlighting the dipolar-octupolar nature of the \nd\ ion \cite{Huang14} have shown that the relevant variable is a pseudo spin $\tau_i$ ($i=x,y,z$ defined in the local ion frame) and which resides on the pyrochlore lattice sites. The magnetic moment ${\bf J}$, hence the observable quantity, essentially identifies with the $\tau^z_i$ component of this pseudo spin and is carried by the local ${\bf z}=\langle 111 \rangle$ axes, consistent with the observed Ising nature of the \nd\ ion. The $x$ and $y$ components identify with octupolar moments of the 4f electronic distribution and are thus barely visible to neutrons \footnote{A weak contribution to the neutron scattering cross section is expected at large wavevectors}, but importantly the $x$ component transforms like a dipole and thus can be coupled to the $z$ component. The low energy properties are then governed by:
\begin{equation*}
{\cal H} = \sum_{\langle i,j \rangle}\left[{\sf J}_{x} \tau^x_i \tau^x_j + {\sf J}_{y} \tau^y_i \tau^y_j + {\sf J}_{z} \tau^z_i \tau^z_j 
+ {\sf J}_{xz} (\tau^x_i \tau^z_j+\tau^z_i \tau^x_j)\right]
 \label{hxyz}
\end{equation*}
which, after a rotation in spin space by an angle $\theta$, can be recast in the XYZ Hamiltonian \cite{Huang14, Benton_2016}:
\begin{equation}
\label{ham}
{\cal H} = \sum_{\langle i,j \rangle} \left[\tilde{{\sf J}}_{\tilde x} \tau^{\tilde{x}}_i \tau^{\tilde{x}}_j + \tilde{{\sf J}}_{\tilde y} \tau^{\tilde{y}}_i \tau^{\tilde{y}}_j + \tilde{{\sf J}}_{\tilde z} \tau^{\tilde{z}}_i \tau^{\tilde{z}}_j \right]
\end{equation}
\begin{equation*}
\begin{aligned}
{\rm with \quad} &\tilde{{\sf J}}_{\tilde x} = \frac{1}{2}\left( {\sf J}_{x}+{\sf J}_{z} + \sqrt{( {\sf J}_{x}-{\sf J}_{z})^2+4{\sf J}_{xz}^2} \right) \\
&\tilde{{\sf J}}_{\tilde z} = \frac{1}{2}\left( {\sf J}_{x}+{\sf J}_{z} - \sqrt{( {\sf J}_{x}-{\sf J}_{z})^2+4{\sf J}_{xz}^2} \right) \\
&\tan 2\theta = \frac{2 {\sf J}_{xz}}{{\sf J}_{x}-{\sf J}_{z}}
\end{aligned}
\end{equation*}
and where the rotation away from the ${\bf z}$ axis defines the new components $(\tau^{\tilde{x}}, \tau^{\tilde{y}}, \tau^{\tilde{z}})$ and the new coupling constants $(\tilde{{\sf J}}_{\tilde x}, \tilde{{\sf J}}_{\tilde y}, \tilde{{\sf J}}_{\tilde z})$. According to the most recent investigations \cite{Xu_2019, Leger20}, the parameters yielding an accurate description of the spin dynamics are $\tilde{{\sf J}}_{\tilde x} \simeq 1, \tilde{{\sf J}}_{\tilde z} \simeq -0.5$~K and  $\tilde{{\sf J}}_{\tilde y}$ close to zero. The corresponding ground state is an AIAO state with the magnetic moments aligned along the ${\bf \tilde z}$ directions, as proposed in Ref. \cite{Benton_2016}. It is worth noting that the ordered pseudo spin is close to its full value $1/2$. The tilted angle $\theta=1-1.2$ rad is estimated from the Curie-Weiss constant, or from the observed ordered moment. Indeed, the projection of the ${\bf \tilde z}$ ordered moment back to the original ${\bf z}$ axes introduces a factor of $\cos \theta$, hence providing a simple explanation for the experimental observation of an AIAO state with a reduced magnetic moment. 

In this work, we combine neutron diffraction and magnetization measurements to explore the evolution of the AIAO magnetic configuration as a function of temperature and magnetic field ${\bf H}$ applied along the three high symmetry directions, and in the presence of non-magnetic Ti substitution at the B-site of the pyrochlore lattice. This study thus envisages the effect of a Zeeman term added to the XYZ Hamiltonian (\ref{ham}):
\begin{equation}
\label{zeeman}
\begin{aligned}
{\cal H}_Z &= \sum_i -g_z \mu_{\rm B}~{\bf \tau^z}_i.{\bf H} \\
&= \sum_i -g_z \mu_{\rm B}
\left[ \cos \theta \tau^ {\tilde{z}}_i + \sin \theta \tau^{\tilde{x}}_i \right] ({\bf z}_i.{\bf H}) 
\end{aligned}
\end{equation}
${\bf z}_i$ are the local $\langle 111 \rangle$ directions and $g_z$ is the anisotropic $g$ factor, estimated to 4.8 in the pure compound and 5 in the 2.5 and 10 \% Ti substituted samples. In addition to classical parameters like Curie-Weiss, effective moment and N\'eel temperature, the detailed analysis of the $M$ vs $H$ curves highlights significant field induced processes, like metamagnetic transitions, that are further analyzed by means of neutron diffraction. This methodology allows one to determine in a systematic way how the microscopic magnetic structure evolution is related to the changes that occur at a more macroscopic scale.

It is worth noting that, as described above, the measured ordered component in zero field is only a fraction of the total magnetic moment of the \nd\ ion ($2.4 - 2.5~\mu_{\rm B}$) (See Table \ref{Table_charac}). Nevertheless, in the presence of a magnetic field, the pseudo spins tilt towards the local ${\bf z}$ directions, so that the measured ordered moment (which is only the ${\bf z}$ component) recovered in high field corresponds to the total magnetic moment.

%%%%%%%%%%%%%%%%%%%%%%%%%%%%
%METHODOLOGY 
%%%%%%%%%%%%%%%%%%%%%%%%%%%%

\section{Methods}

\subsection{Samples}
\ndzrx\ samples (with $x = 0.025$ and $0.1$) were prepared at the University of Warwick in polycrystalline form by standard solid-state chemistry methods, using Nd$_2$O$_3$, ZrO$_2$ and TiO$_2$ as starting reagents \cite{Ciomaga_2015, Ciomaga_2016}. To ensure the appropriate composition of the final compounds, the Nd$_2$O$_3$ powder was pre-heated in air at $1000^\circ$C for 24 hours. Powders of the starting oxides were weighed in stoichiometric amounts, mixed together, pressed into pellets and then heat treated in air for several days (in three or fours steps) at temperatures in the range $1400-1500^{\circ}$C. The annealed pellets were reground between each step of the synthesis to ensure good homogeneity and to facilitate the chemical reaction. Powder X-ray diffraction measurements confirmed the phase purity of the synthesized \ndzrx\ polycrystalline samples.

Single crystals of the two Ti-substituted samples were grown from the polycrystalline samples by the floating-zone technique using a four-mirror xenon arc lamp optical image furnace \cite{Ciomaga_2015,Ciomaga_2016}. The 2.5 \% substituted sample is a fragment of the single crystal used in Ref. \onlinecite{Leger20}. 

The pure sample is the same as in Ref. \onlinecite{Lhotel_2015, Petit_2016}.

Below, we report only on the single crystal properties. Powder sample properties are summarized in Appendix~\ref{powder}.

\subsection{Bulk magnetic measurements}

Magnetization and AC susceptibility measurements were performed down to 80~mK on the single crystal samples using superconducting quantum interference device (SQUID) magnetometers, equipped with a dilution refrigerator developed at the Institut N\'eel-CNRS Grenoble \cite{Paulsen01}. The crystals were glued with GE varnish on a copper plate, and measurements were collected with the magnetic field aligned along the three high symmetry directions of the crystals ($[111]$, $[110]$ or $[001]$). All the data shown below were corrected from demagnetizing field effects. Demagnetizing factors were estimated from the sample shapes using the Aharoni formula \cite{Aharoni98}. 

\subsection{Neutron diffraction}
Single crystal diffraction experiments were carried out at D23 (CRG CEA-ILL, France) using a wavelength $\lambda = 1.275$ \AA\ and operated with a copper monochromator. The samples were glued on the Cu finger of a dilution insert and placed in a cryomagnet. The experiments were conducted with the (vertical) field ${\bf H}$ either parallel to the $[001]$, $[1\bar{1}0]$ or $[111]$ high symmetry crystallographic direction. The three orientations were measured for the three samples, except for the 10 \% substituted sample for which the ${\bf H} \parallel [1\bar{1}0]$ condition could not be measured.

\begin{table}[!]
\begin{tabularx}{7cm}{c*{6}{Y}}
\hline
\hline
\multirow{2}{*}{Nd Atoms} & \multirow{2}{*}{$X$} & \multirow{2}{*}{$Y$} & \multirow{2}{*}{$Z$} & \multirow{2}{*}{$n_x$} & \multirow{2}{*}{$n_y$} & \multirow{2}{*}{$n_z$} \\
\\
\hline
%1
Nd1 & 0.25 & 0.25 & 0.50 & 1 & 1 & -1 \\
%2
Nd2 & 0.00 & 0.00 & 0.50 & -1 & -1 & -1 \\
%3
Nd3 & 0.00 & 0.25 & 0.75 & -1 & 1 & 1 \\
%4
Nd4 & 0.25 & 0.00 & 0.75 & 1 & -1 & 1 \\
\hline
\hline
\end{tabularx}
\caption{\label{table_ND} 
Sign convention used in the {\sc FullProf} refinements. Atoms listed as Nd1 to Nd4 (with $(X, Y, Z)$ positions) form the outer tetrahedron sketched in red in Figure~\ref{Image1}. With this convention, positive values of the magnetic moments correspond to outgoing spins, parallel to the Ising vector ${\bf z} = (n_x, n_y, n_z)$. The twelve other atoms are obtained by symmetry operations.}
\end{table}

As we shall see later on, the magnetic structures have a ${\bf K}= {\bf 0}$ propagation vector, hence magnetic and crystalline intensities occur on the same $(h,k,l)$ positions. A full data collection was thus first measured at $6$ K to serve as a reference of the crystalline intensities and refine the sample volume, the Ti content $x$, the extinction parameters and atomic positions in the Fd$\bar{3}$m space group. Additional data collections (consisting in 80 Bragg peaks) were then measured at low temperature (80 mK), at selected fields ranging from $-1$ T to $+1$ T. The magnetic structure at each field was then refined with the {\sc Fullprof} suite \cite{Fullprof} using the difference between the low temperature and 6~K diffractogram intensities. Owing to the strong Ising character of the \nd\ ions, the data have been analyzed assuming that the magnetic moments lie along the ${\bf z}$ Ising axes (see Table \ref{table_ND}). The sign, with respect to the convention given in Table \ref{table_ND} and the amplitude of the moments in a given tetrahedron are the free parameters of those fits. 

In addition, the intensity of selected Bragg peaks was recorded while ramping the field back and forth from $-1$~T to $+1$ T ($0.015$~T/min was the minimum sweeping speed) to obtain a ``continuous'' evolution vs field. The sample was first cooled down to 80 mK in zero field and the field dependence of the intensity was then measured after the application of a $+1$ T field to saturate the sample (larger fields up to $+3$ T were also used, but this makes no changes in the ``low" field results). The detailed method for the analysis of these field sweeping measurements is given in Appendix \ref{ramp_analysis}.

\begin{table*}[t!]
\begin{tabularx}{\textwidth}{cYcY*{4}{c}}
\hline
\hline
\multirow{2}{*}{Single crystal Sample} & Ti concentration & \multirow{2}{*}{O2 position} & Lattice parameter & $T_{\rm N}$ & $\theta _{\rm CW}$ & $\mu _{\rm eff}$ & Ordered moment \\
& [\%] && [\AA] & [mK] & [mK] & [$\mu _{\rm B}$] & [$\mu _{\rm B}$] \\
\hline
%1
$\mathrm{Nd_2Zr_2O_7}$            & 0   & $0.337 \pm 0.002$  & $10.66 \pm 0.02$ & 285 & 195 & 2.45 & $0.8 \pm 0.05$ \\
%2
$\mathrm{Nd_2Zr_{1.95}Ti_{0.05}O_7}$ & $2.41\pm0.2$ & $0.336 \pm 0.001$ & $10.65 \pm 0.02$ & 375 & 235 & 2.52 &$1.19 \pm 0.03$\\
%3
$\mathrm{Nd_2Zr_{1.8}Ti_{0.2}O_7}$   & $7.9\pm0.7$  & $0.336 \pm 0.001$ & $10.64 \pm 0.03$ & 325 & 220& 2.50 & $1.06 \pm 0.04$ \\
\hline
\hline
\end{tabularx}
\caption{\label{Table_charac} Sample (single crystals) parameters obtained from the structural and magnetic {\sc Fullprof} refinements at 6 K and from magnetization measurements. The results from the refinements are averaged over the three sets of measurements corresponding to different scattering planes. %The values for the pure compound are from Refs. \onlinecite{Ciomaga_2015,Lhotel_2015}.
}
\end{table*}

%%%%%%%%%%%%%%%%%%%%%%%%%%%%%%%
%RESULTS/CARACTERISATIONS 
%%%%%%%%%%%%%%%%%%%%%%%%%%%%%%%
\section{Results}

\subsection{Physical characterizations}

\begin{figure}[h!]
\includegraphics[width=7cm]{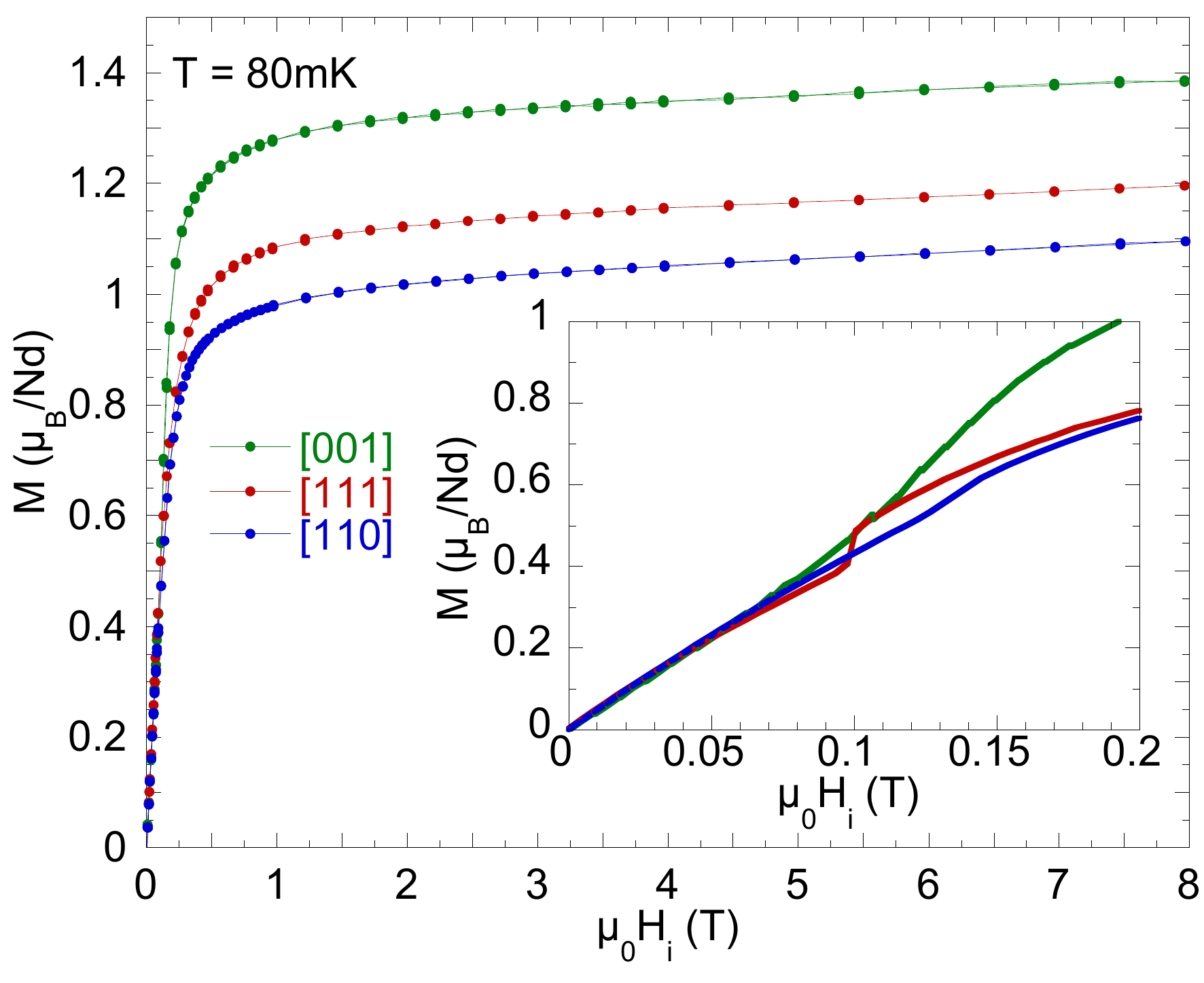}
\caption{\label{MH} 
Magnetization $M$ versus internal field $H_i$ measured with the field applied along the three main high symmetry directions at 80 mK on the 10 \%-Ti substituted sample $\mathrm{Nd_2Zr_{1.8}Ti_{0.2}O_7}$. Similar curves are observed in the 2.5 \%-Ti substituted sample. Inset : Zoom in the low field data.}
\end{figure}

The structural refinements from neutron diffraction data of the single crystals with the 2.5 and 10 \% substitution concentration are in agreement with the pyrochlore structure. The lattice parameters are close to the pure compound one ($10.66$ \AA) and the 48f oxygen atoms are found at the position $x_{\rm 48f} = 0.336$. Noteworthy, the value of the Ti content was refined thanks to the significant difference in the scattering length for the two elements Zr and Ti. It is found to be 2.4 and 7.9 \% in the so-called 2.5 and 10 \% Ti-substituted samples respectively (see Table \ref{Table_charac}).

Isothermal magnetization curves for magnetic fields applied along the three high symmetry directions show three different saturated magnetizations (see Figure \ref{MH}), in agreement with expectations in the case of $\langle 111 \rangle$ Ising spins located at the vertices of the tetrahedra \cite{Harris_1998}. Curie-Weiss fits of the magnetization between $1$ and $4$ K give positive Curie-Weiss temperatures $\theta _{\rm CW}$ and effective moments of about $2.5~\mu_{\rm B}$ for both Ti-substituted samples.

\begin{figure}[t!]
\includegraphics[width=8cm]{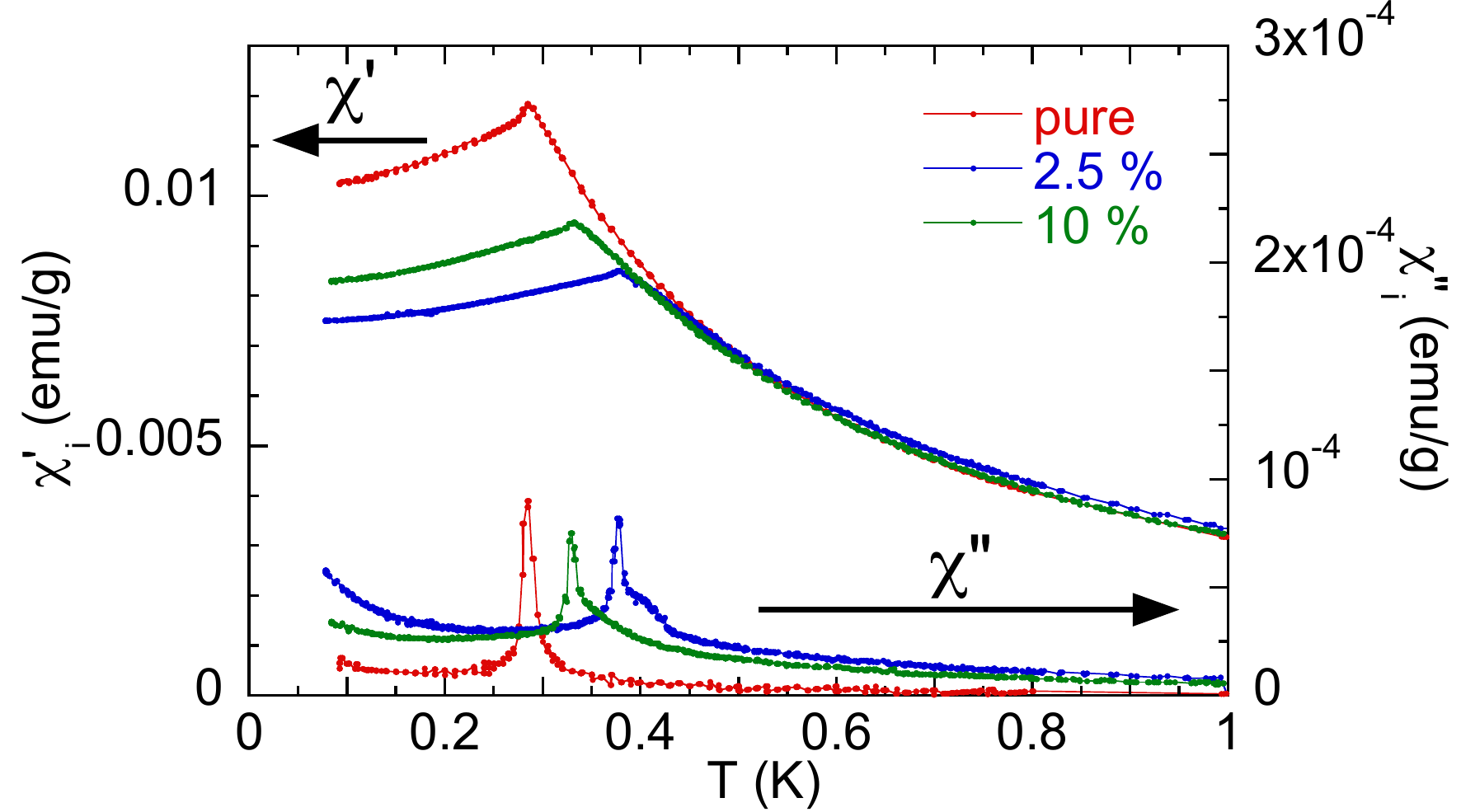}
\caption{\label{Xac} 
Real and imaginary parts of the AC susceptibility $\chi'$ (left) and $\chi''$ (right) for the pure (red dots), 2.5 \% (blue dots) and 10\% (green dots) substituted samples, along the $[1\bar{1}0]$ direction. Measurements were performed with $H_{\rm AC}=2.7$ Oe and a frequency $f=5.7$ Hz. }
\end{figure}

AC susceptibility measurements show a transition at low temperature in all compounds with a peak in both real and imaginary parts (see Figure \ref{Xac}). No frequency dependence of the peak position is observed between 0.057 and 570 Hz. Interestingly, the critical temperature $T_{\rm N}$ reaches 376~mK for $x=2.5$ \% and decreases back to 325~mK for $x=10$ \%. Neutron diffraction measurements elucidate that the magnetic ground state is AIAO in the substituted samples with $1.2$ and $1.1~\mu_{\rm B}$ for the 2.5 and 10~\% substituted samples respectively. For both samples, the transition takes place at a higher temperature than in the pure compound ($T_{\rm N}=285$ mK) and with an ordered magnetic moment significantly larger (0.8 $\mu_{\rm B}$ in the pure compound), as summarized in Table \ref{Table_charac}. It should be noted that these \TN\ and ordered moment values for the substituted samples are nevertheless smaller than the ones measured on the powder samples with the same composition (See Appendix \ref{powder}).

%AC susceptibility measurements show a transition at low temperature in all compounds with a peak in both real and imaginary parts (see Figure \ref{Xac}). No frequency dependence of the peak position is observed between 0.057 and 570 Hz. Interestingly, the critical temperature $T_{\rm N}$ increases up to $376$ mK for $x=2.5$ \% and decreases back to $325$ mK for $x=10$ \%. In both substituted samples, the transition takes place at a higher temperature than in the pure compound ($T_{\rm N}=285$ mK), but at lower temperature than in the corresponding powder samples (See Appendix \ref{powder}). Neutron diffraction measurements elucidate that the magnetic ground state is AIAO in the substituted samples, with an ordered magnetic moment significantly larger that in the pure sample, with $1.2$ and $1.1~\mu_{\rm B}$ for the 2.5 and 10 \% substituted samples respectively. Here again, these values are slightly smaller than in the powder samples. These results are summarized for the three samples in Table \ref{Table_charac}.

At low field and in all samples, the $M$ vs $H$ curves show an inflexion point (see the insert in Figure \ref{MH}) for the three field directions, which is associated to metamagnetic like processes \cite{Lhotel_2015}. The corresponding characteristic field can be followed by tracking the maximum of the derivative $dM/dH_i$ vs $H$ of these magnetization curves for different temperatures in order to build the $(H,T)$ phase diagram. In the following sections, the nature of the phases is investigated based on neutron diffraction data.

%%%%%%%%%%%%%%%%%%%%%%%%%
%RESULTATS/001 
%%%%%%%%%%%%%%%%%%%%%%%%%

\subsection{Magnetic field along $[001]$}
\label{001}
\begin{figure*}[!t]
\includegraphics[width=17cm]{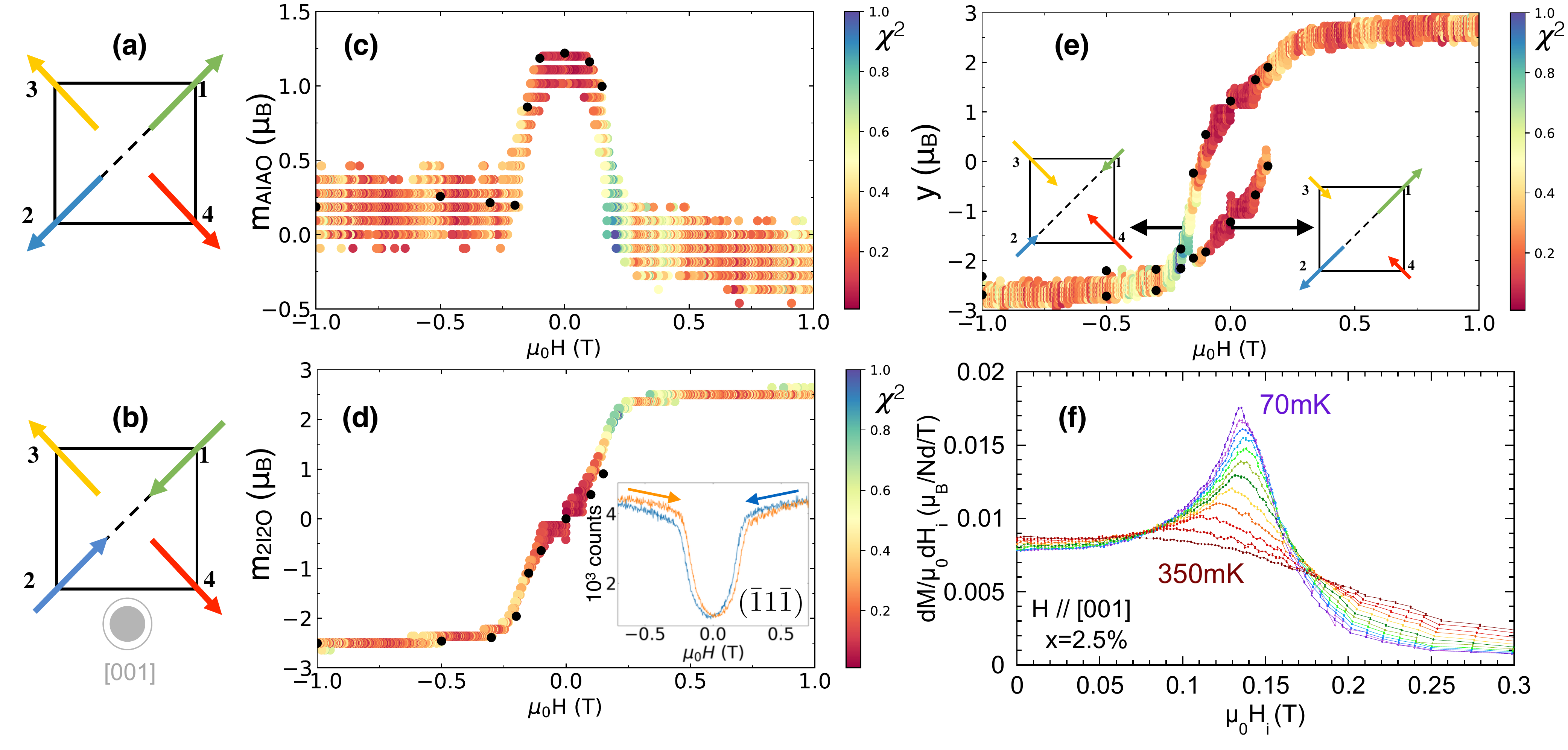}
\caption{\label{res001} 
Field dependence for ${\bf H} \parallel [001]$ in the 2.5\%-substituted sample. (a) Sketch of the 4 spins in the ``all out'' state, obtained in zero field and (b) of the 2I2O state obtained at saturation. (c-d-e) Neutron diffraction results: (c) $m_{\rm AIAO}$, (d) $m_{\rm 2I2O}$ and (e) total $y$ component versus $H$ (see Equation \ref{m_001}). The left (respectively right) branch results from $m_{\rm AIAO}>0$ - ``all out" component ($m_{\rm AIAO}<0$ - ``all in" component). {\sc Fullprof} refinements are displayed as full black dots. Colored points indicate the most probable values obtained from the field sweeping measurements analysis (see Appendix \ref{ramp_analysis}). The color scale indicates the $\chi^2$ value obtained from Equation \ref{chi2}. The inset of (d) shows the intensity of the $(\bar{1}1\bar{1})$ reflection. (f) $dM/dH_i$ versus $H_i$. The field was always swept from negative to positive values.}
\end{figure*}

When applying a magnetic field along the $[001]$ direction, the AIAO ground state is expected to evolve towards an ordered spin ice state, with two incoming and two outgoing spins per tetrahedron and a net magnetization along the $[001]$ direction. According to symmetry, the 4 spins in a given tetrahedron form two subgroups consisting in spins (1,2) and (3,4) (see Figure \ref{Image1}). To follow the magnetic structure and figure out how the ordered spin ice configuration grows to the detriment of the ``all in / all out" one, it is thus convenient to parametrize the magnetic moments as:
\begin{equation}
\label{m_001}
\begin{array}{lcl}
{\bf m}_1 &=& (m_{\rm AIAO}-m_{\rm 2I2O})~{\bf z}_1 = x~ {\bf z}_1 \\
{\bf m}_2 &=& (m_{\rm AIAO}-m_{\rm 2I2O})~{\bf z}_2 = x~ {\bf z}_2 \\
{\bf m}_3 &=& (m_{\rm AIAO}+m_{\rm 2I2O})~{\bf z}_3 = y~ {\bf z}_3 \\
{\bf m}_4 &=& (m_{\rm AIAO}+m_{\rm 2I2O})~{\bf z}_4= y~{\bf z}_4 
\end{array}
\end{equation}
which corresponds to the sum of a generalized ``all in / all out" component $m_{\rm AIAO}$ and of a ``2 in -- 2 out" component $m_{\rm 2I2O}$. By convention, considering a red tetrahedron of Figure \ref{Image1}, $m_{\rm AIAO}$ is positive (negative) when all the Nd magnetic moments are ``out" (``in"), and $m_{\rm 2I2O}$ is positive when Nd1 and Nd2 are ``in'' while Nd3 and Nd4 are ``out'' (2I2O configuration) and negative when  Nd1 and Nd2 are ``out'' while Nd3 and Nd4 are ``in'' (2O2I configuration).

The {\sc Fullprof} refinements highlight this field induced behavior, as shown in Figure \ref{res001} for the 2.5 \% substituted sample. Starting from $\mu_0 H=-1$ T, red tetrahedra of Figure \ref{Image1} are in the 2O2I configuration with a net moment along $[0 0 {\textrm -1}]$. Upon increasing field, this ``2 in -- 2 out" component remains stable (decreasing smoothly) up to a field threshold from which it is suppressed to the benefit of the AIAO component, as shown in the panels (c) and (d) of Figure~\ref{res001}. This characteristic field is clearly observed when measuring the intensity of the $(\bar{1} 1 \bar{1})$ reflection as a function of field, as shown in the inset of Figure \ref{res001}(d).

The AIAO component is maximum in zero field. It decreases when further increasing the field, while $m_{\rm 2I2O}$ rises continuously, until reaching a value close to the saturation at the opposite of the first threshold field, i.e. when the AIAO component comes back to zero. This clearly separates two regions in the phase diagram: above a characteristic field $H_{001}$, only a 2I2O component is present, while below, it coexists with an AIAO one.  

\begin{figure}[t!]
\includegraphics[width=8cm]{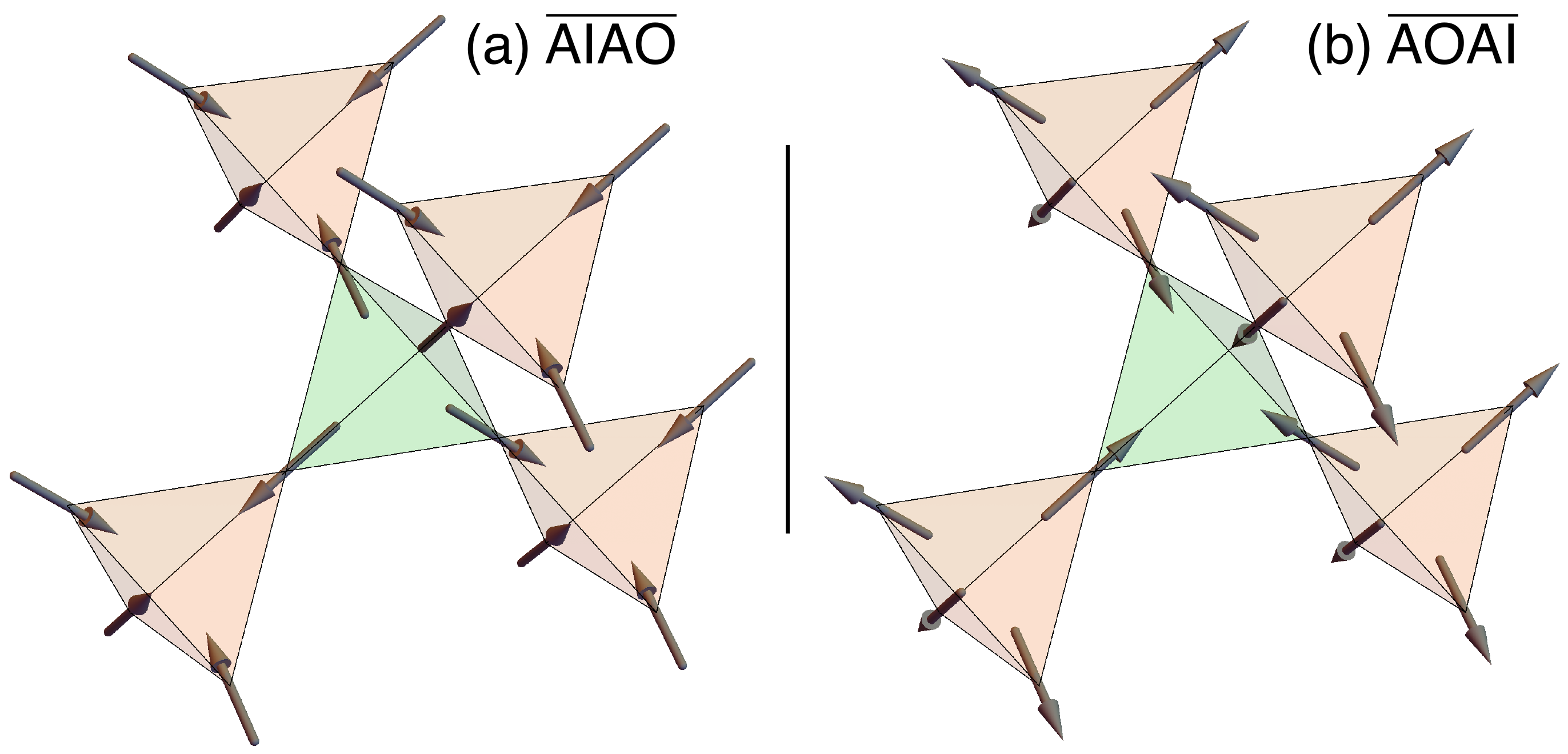}
\caption{\label{domains} 
Pictures of the two domains of the AIAO state: (a) ${\rm \overline{AIAO}}$ domain: red tetrahedra are ``all in'' and green tetrahedra are ``all out". (b) ${\rm \overline{AOAI}}$ domain: red tetrahedra are ``all out'' and green tetrahedra are ``all in". }
\end{figure}

The analysis nevertheless points to two different situations upon increasing field from the 2O2I configuration: (i) either Nd1 and Nd2 flip, so that the Figure \ref{Image1} red tetrahedra become ``all in'' (and green tetrahedra become ``all out"), or (ii) Nd3 and Nd4 flip, so that the red tetrahedra become ``all out''. These two magnetization processes lead to the formation of the two domains of the AIAO state, so-called ${\rm \overline{AIAO}}$ or ${\rm \overline{AOAI}}$ (See Figure~\ref{domains}), which cannot be distinguished in neutron diffraction measurements. The two domains have the same energy in the presence of the $[001]$ field which suggests that both are present simultaneously. Interestingly, the resulting moment ($x$ or $y$ in Equation \ref{m_001}) on a given atom is different depending on the sign of $m_{\rm AIAO}$, and thus on which ${\rm \overline{AOAI}}$ domain is stabilized. This manifests through the two branches shown in the panel (e) of Figure \ref{res001} representing the magnetic moment of the Nd3 atom. Finally, almost no hysteresis is observed in magnetization and neutron field sweeping measurements, which indicates that both domains are present and that the magnetization process takes place while remaining close to thermodynamic equilibrium.

The signature of $H_{001}$ in the magnetization curves vs field is an inflexion point, which appears as a broad peak in the $dM/dH_i$ vs field curves (see Figure \ref{res001}(f)). At the lowest temperature (about 80 mK), $H_{001}$, determined from the maximum of $dM/dH_i$, varies from $0.14$ up to $0.25$ T depending on the sample composition (see Table~\ref{Table_HM}). The values of $H_{001}$ as a function of temperature are plotted for the three samples on the $(H,T)$ phase diagram of Figure \ref{resHT}(a). As expected, $H_{001}$ goes to zero at the N\'eel temperature, confirming that it is characteristic of the disappearance of the AIAO structure. The temperature dependence of $H_{001}$ is qualitatively similar in the three samples. Interestingly, the lower the $T_{\rm N}$, the lower the $H_{001}$. Consequently, the pure compound has the lowest $H_{001}$ and the 2.5 \% Ti-substituted sample has the highest one. 

%%%%%%%%%%%%%%%%%%%%%%%%%
%RESULTATS/110 
%%%%%%%%%%%%%%%%%%%%%%%%%
\subsection{Magnetic field along $[1\bar{1}0]$}
\label{110}
\begin{figure*}[!t]
\includegraphics[width=17cm]{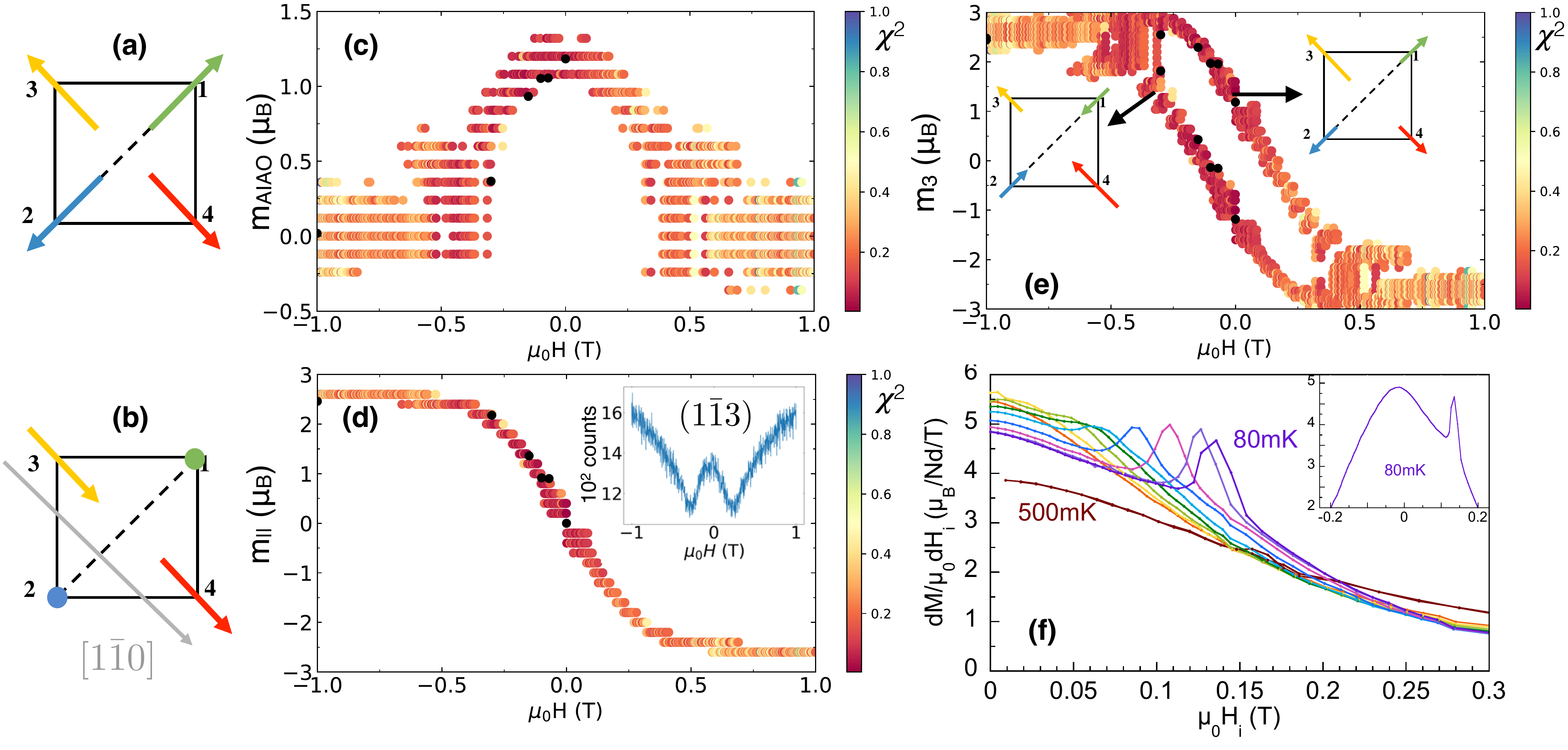}
\caption{\label{res110} 
Field dependence for ${\bf H} \parallel [1\bar{1}0]$ in the 2.5\%-substituted sample. (a) Sketch of the 4 spins in the ``all out'' state, obtained in zero field and (b) of the partially ordered state obtained at saturation.
(c-d-e) Neutron diffraction results: (c) $m_{\rm AIAO}$, (d) $m_{\parallel}$ and (e) total $m_3$ component versus $H$ (see Equation \ref{m_110}). The right (respectively left) branch results from $m_{\rm AIAO}>0$ - ``all out" component ($m_{\rm AIAO}<0$ - ``all in" component). {\sc Fullprof} refinements are displayed as full black dots. Colored points indicate the most probable values obtained from the field sweeping measurements analysis (the color scale corresponds to the $\chi^2$ value, see Appendix \ref{ramp_analysis}). The inset of (d) shows the intensity of the $(1\bar{1}3)$ reflection. (f) $dM/dH_i$ versus $H_i$ at different temperatures between 80 and 500 mK. The inset shows the whole curve at 80 mK. The field was always swept from negative to positive values.}
\end{figure*}

In the presence of a $[1\bar{1}0]$ field, the situation is more complicated. In such a field direction, the system is usually described as two perpendicular chains, made of the spins 1 and 2 ($\beta$ chains), and spins 3 and 4 respectively in our notations ($\alpha$ chain, see Figure \ref{Image1}). Spins 3 and 4 are almost collinear to the field and are thus expected to quickly reach saturation. In contrast, spins 1 and 2 have their Ising direction perpendicular to the field. The Zeeman effect is thus zero, yet the field indirectly affects their orientation via the coupling to the other spins \cite{Harris_1998}. To describe the magnetic structure as a function of field, the following parametrization was thus used:
\begin{equation}
\label{m_110}
\begin{array}{lcl}
{\bf m}_1 &=& (m_{\rm AIAO}+m_{\perp})~{\bf z}_1 \\
{\bf m}_2 &=& (m_{\rm AIAO}-m_{\perp})~{\bf z}_2 \\
{\bf m}_3 &=& (m_{\rm AIAO}+m_{\parallel})~{\bf z}_3 \\
{\bf m}_4 &=& (m_{\rm AIAO}-m_{\parallel})~{\bf z}_4
\end{array}
\end{equation}
It takes into account the ``all in / all out" contribution $m_{\rm AIAO}$ (with the same sign convention as before) along with $m_{\parallel}$ and $m_{\perp}$ which feature field polarized components parallel (for spins 3 and 4) and perpendicular to the nominal field direction (for spins 1 and 2).

{\sc Fullprof} refinements of the neutron diffraction measurements highlight the saturation of the parallel spins in large magnetic fields: at $|\mu_0H|=1$ T, the Nd3 and Nd4 moments reach $2.4-2.5~\mu_{\rm B}$, the Nd3 moment being ``in'', and the Nd4 one ``out'' in positive field (see Figure~\ref{res110}(b)), and conversely in negative field. Noteworthy, they reach values as high as $1.5~\mu_{\rm B}$ at $|\mu_0H|=0.15$ T (see Figure~\ref{res110}(d)). At high field, the refined magnetic moment on sites 1 and 2 is found to be zero (or small \footnote{It is worth mentioning that in a trial experiment where the field was not perfectly aligned along $[1\bar{1}0]$ (about 4 degrees while less than 1 in the present experiment), such a non-zero moment could be refined, suggesting that it is likely a by product of the ill-aligned component of the field}) (see Figure \ref{res110}(c)). This confirms the disordered character of the (Nd1, Nd2) chains in applied field previously observed in Ref. \onlinecite{Xu_2018}. This study further reports diffuse scattering showing the existence of short-range correlations within the $\beta$ chains, that we could not address in the present work. 

Neutron refinements (see Figure \ref{res110}c) show that the AIAO component is suppressed upon increasing the field (in absolute values) at a critical value labeled $H_{110}$. It is best observed when measuring the characteristic $(1 \bar{1} 3)$ magnetic reflection as a function of the field, as shown in the inset of Figure \ref{res110}(c).

Like in the case of the $[001]$ field, the analysis of the field sweeping measurements unveils two scenarii regarding the rise of the ``all in / all out" component when starting from the high field state, and corresponding to the possible ${\rm \overline{AIAO}}$ and ${\rm \overline{AOAI}}$ domains. Indeed, considering the (Nd3, Nd4) $\alpha$ chain in the saturated state, either the Nd3 or the Nd4 spin can flip to recover ``all in" or ``all out" configurations, with the same energy cost. It results in two possible branches in the total magnetic moment, as shown for the Nd3 atom in Figure \ref{res110}(e).

%Like in the case of the $[001]$ field, the ramp analysis unveils two scenarii regarding the rise of the ``all in / all out" component when starting from the high field state, and corresponding to the possible AIAO and AOAI domains. Indeed, considering the (Nd3, Nd4) $\alpha$ chain (see Figure \ref{res110}(b)), either the Nd3 or the Nd4 spin can flip to recover an AIAO state, leading to either ``all in" or ``all out" configurations for the red tetrahedra of Figure~\ref{Image1}. 

The signature of $H_{110}$ in the magnetization curves vs field is characterized by a step-like anomaly and manifests as a little peak in the $dM/dH_i$ curves (see Figure~\ref{res110}(d)), which is more marked in the substituted samples than in the pure sample. Nevertheless, this peak is only clearly observed in positive field, when sweeping the field from negative values, as shown in the inset of Figure~\ref{res110}(d) at 80 mK. A small irreversibility is also observed on the $(2 2 0)$ reflection (not shown) in neutron diffraction measurements. These results imply that some kind of metastable states exist for this field direction, that we will address in more details in the discussion. The temperature dependence of $H_{110}$ is qualitatively similar in the three samples, and as for the $[001]$ direction, the pure compound has the lowest $H_{110}$ while the 2.5 \% Ti-substituted sample has the highest one. 

%%%%%%%%%%%%%%%%%%%%%%%%%%%
%RESULTATS/111 
%%%%%%%%%%%%%%%%%%%%%%%%%%%
\subsection{Magnetic field along $[111]$}
\label{111}
\begin{figure*}[!t]
\includegraphics[width=17cm]{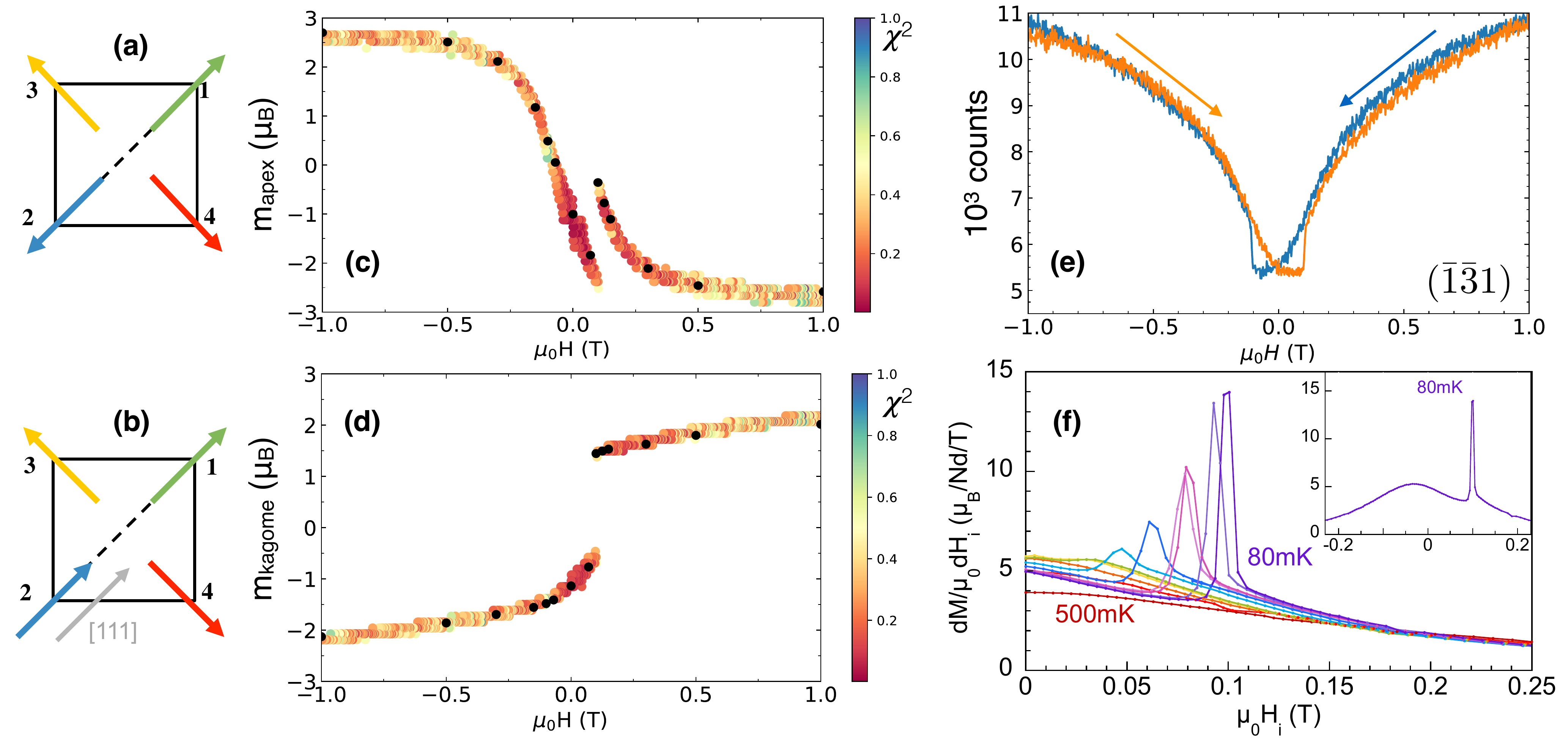}
\caption{\label{res111} 
Field dependence for ${\bf H} \parallel [111]$ in the 10\%-substituted sample. (a) Sketch of the 4 spins in the ``all out'' state, obtained in zero field and (b) of the 1I3O state obtained at saturation in a positive field.
(c-e) Neutron diffraction results: (c) $m_{\rm apex}$ and (d) $m_{\rm kagome}$ components versus $H$ (see Equation \ref{m_111}). {\sc Fullprof} refinements are displayed as full black dots. Colored points indicate the most probable values obtained from the field sweeping measurements analysis (the color scale corresponds to the $\chi^2$ value, see Appendix \ref{ramp_analysis}). (e) Intensity of the $(\bar{1}\bar{3}1)$ reflection versus $H$. (f) $dM/dH_i$ versus $H_i$ at different temperatures between 80 and 500 mK. The inset shows the whole curve at 80 mK. The field was always swept from negative to positive values.}
\end{figure*}

To discuss the case of a field applied along the $[111]$ direction, it is convenient to view the pyrochlore lattice as the stacking of triangular and kagom\'e planes. In a given tetrahedron, one shall then distinguish the apex spin, which is collinear to the field (Nd2), and three ``kagom\'e'' spins located in the kagom\'e layer (Nd1, 3 and 4), which experience the same Zeeman effect. The magnetic configuration is expected to go from a 1O3I to ``all in" or ``all out", and eventually to 1I3O with increasing the field from $-1$ up to $+1$ T, a behavior described in the pure compound in Ref. \cite{Lhotel_2018, Opherden_2017}. To describe this evolution, the following parametrization is used: 
\begin{equation}
\label{m_111}
\begin{array}{lcl}
{\bf m}_1 &=& m_{\rm kagome}~{\bf z}_1 \\
{\bf m}_2 &=& m_{\rm apex}~{\bf z}_2 \\
{\bf m}_3 &=& m_{\rm kagome}~{\bf z}_3 \\
{\bf m}_4 &=& m_{\rm kagome}~{\bf z}_4
\end{array}
\end{equation}
Neutron diffraction refinements reveal in both Ti-substituted samples the same singularity as the one already observed in the pure compound. It is characterized by a sudden jump for both the apex and the kagom\'e spins (see Figure~\ref{res111}(c,d)). At $\mu_0H=-1$ T, the kagom\'e spins are ``in'' while the apex is ``out'', forming a 1O3I configuration. Upon increasing the field from $-1$ T, the absolute value of the apex moment progressively decreases while the kagom\'e moments remain essentially unchanged. At some threshold $\mu_0H \approx -0.02$ T, the apex moment changes sign, and the structure becomes ``all in''. This holds up to a critical field $\mu_0 H_{111} \approx 0.1$ T where the kagom\'e moments abruptly flip while the apex moment nearly goes back to zero. Its amplitude increases then, but now follows a new magnetization process. The configuration has become 1I3O. The $H_{111}$ field is best observed when measuring the characteristic $(\bar{1}\bar{3}1)$ magnetic reflection as a function of the field, as shown in Figure \ref{res111}(e). 

The sharp transition at $H_{111}$ underlines the existence of metastable states when increasing the field, and which are again connected to the formation of either ``all in" or ``all out" configurations for the red tetrahedra, and thus to ${\rm \overline{AIAO}}$ and ${\rm \overline{AOAI}}$ domains. These domains are connected to the ${\rm \overline{1O3I / 1I3O }}$ and ${\rm  \overline{1I3O / 1O3I}}$ configurations stabilized at large negative and positive fields respectively. Depending on the amplitude of the field, they become metastable, explaining the origin of the hysteresis clearly observed in the magnetization measurements displayed in the inset of Figure \ref{res111}(f). 

\subsection{Phase diagrams}
\begin{table}[!]
\begin{tabularx}{8cm}{*{4}{Y}}
\hline
\hline
Sample & $\mu_0 H_{001}$ (T) & $\mu_0 H_{110}$ (T) & $\mu_0 H_{111}$ (T) \\
\hline
pure & 0.083 & 0.086 & 0.075 \\
2.5 \% & 0.135 & 0.162 & 0.111 \\
10 \% & 0.124 & 0.132 & 0.099 \\
\hline
Calculations &&&\\
%pure & 0.158 & 0.173 & 1.355 \\
%2.5 \% & 0.293 & 0.245 & 1.303 \\
pure & 0.26 & 0.55 & 1.20 \\
2.5 \% & 0.32 & 0.72 & 1.13 \\
\hline
\hline
\end{tabularx}
\caption{\label{Table_HM} 
Critical fields at the lowest temperature for the three symmetry directions, obtained experimentally for the three samples, and by mean-field caluclations for the pure and 2.5 \% samples.}
\end{table}

\begin{figure*}[t!]
\includegraphics[width=\textwidth]{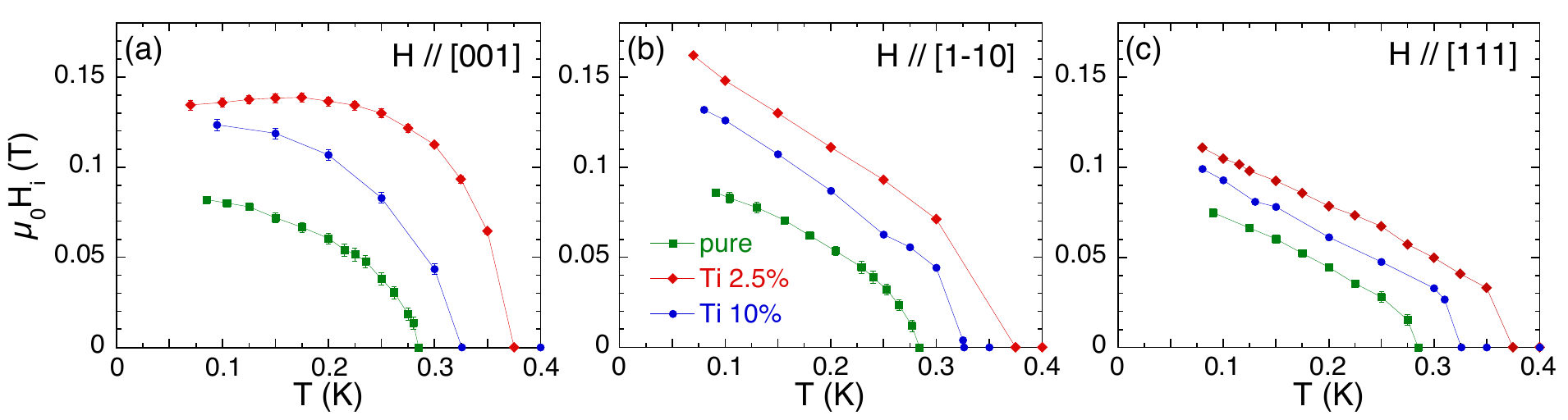}
\caption{\label{resHT} Field - temperature $(H, T)$ phase diagrams determined from $M$ vs $H$ experiments for the three high symmetry directions (a) $[001]$, (b) $[1\bar{1}0]$, (c) $[111]$ and for the studied samples: the pure (green squares), the 2.5 \%-substituted (red diamonds), and the 10 \%-substituted (blue dots) compounds. Lines are guides to the eye.
}
\end{figure*}
\begin{figure*}[!]
\includegraphics[width=\textwidth]{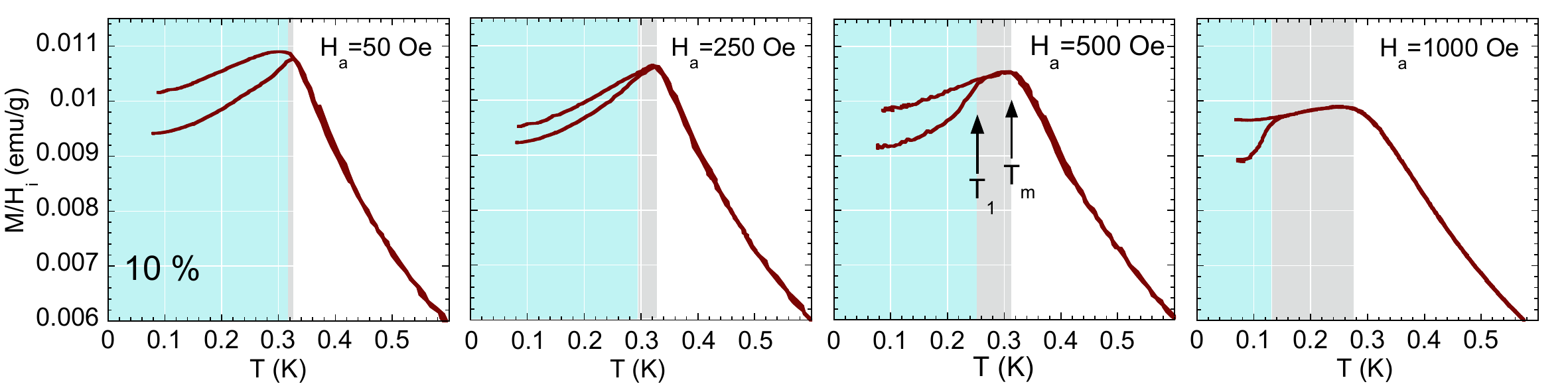}
\caption{\label{ZFC_111} $M/H_i$ versus temperature for the 10\%-Ti substituted sample obtained in ZFC-FC measurements at different applied magnetic fields along $[111]$. $T_1$ is defined as the temperature where the ZFC and the FC curves split and $T_{\rm m}$ is defined as the temperature where $M/H_i$ reaches its maximum.}
\end{figure*}

\begin{figure}[!]
\includegraphics[width=8.5cm]{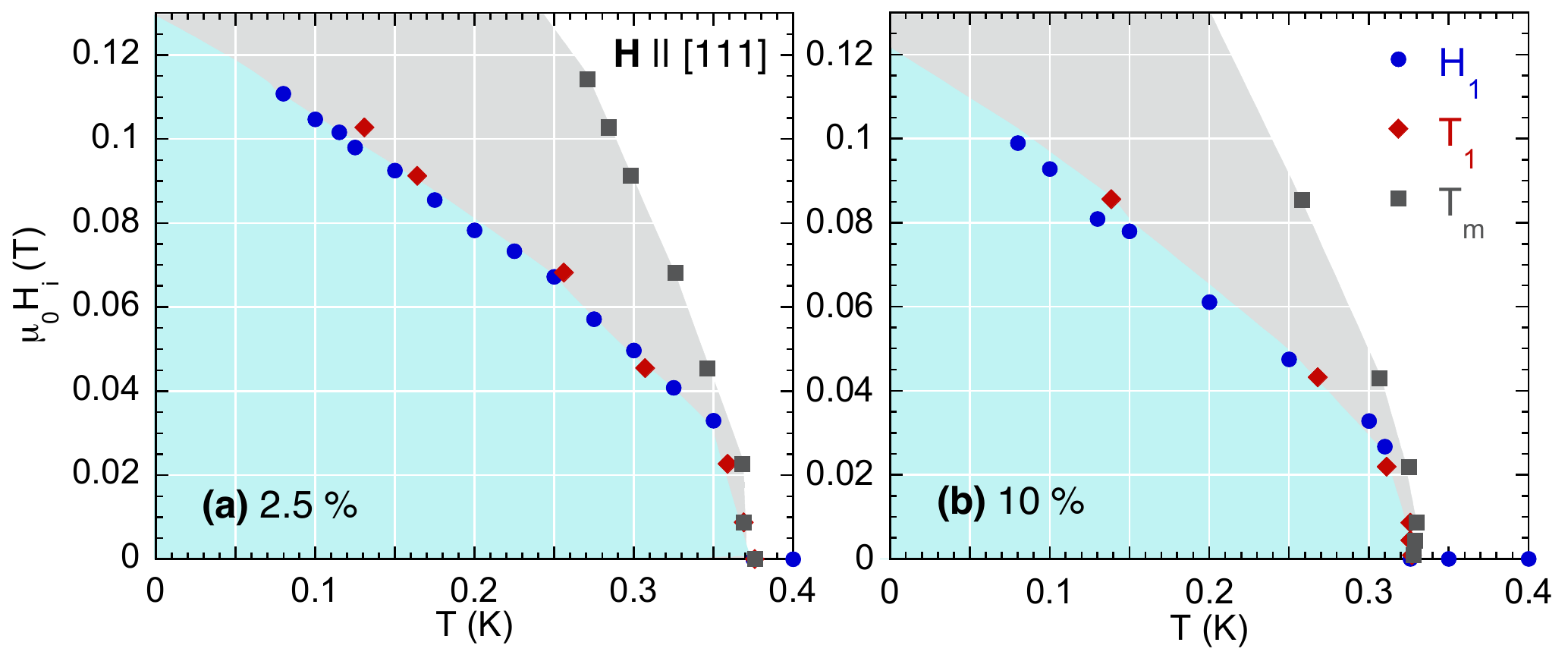}
\caption{\label{fig_HT_111} Phase diagrams $(H,T)$ for ${\bf H} \parallel [111]$ from $M$ vs $T$ ($T_1$, red diamonds, and $T_{\rm m}$, green squares) and $M$ vs $H$ ($H_1$, blue dots) measurements for the 2.5 and 10 \%-substituted samples.}
\end{figure}

From the temperature dependence of the peak in the $dM/dH_i$ vs $H_i$ curves shown in the panels (f) of Figures~\ref{res001}, \ref{res110} and \ref{res111}, we could construct the phase diagrams for the three field directions, and the three measured samples. They are displayed in Figure \ref{resHT}. The three samples show qualitatively the same behavior, the critical fields being related to the critical temperatures in all directions so that the larger the N\'eel temperature, the larger the critical field. 

Along the $[001]$ and $[1\bar{1}0]$ directions, the low field phase can be directly associated to an ``all in / all out" phase, as demonstrated by the neutron diffraction results detailed in the above sections. Along $[001]$, the critical fields seem to saturate at low temperature (and even to decrease slightly for the 2.5 \% sample). On the contrary, the low temperature dependence of the critical field along $[1\bar{1}0]$ does not show any sign of saturation. 

The description in terms of AIAO component for the field along the $[111]$ direction is less relevant due to the different response of the apical and kagom\'e spins with respect to the field in that case. In addition, the observation of a very strong hysteresis suggests that the value of $H_{111}$ is unlikely an equilibrium value, as discussed in more details in Section \ref{discussion}. 

To further probe the nature of the field induced behavior in this field direction, we have performed $M$ vs $T$ measurements in various applied fields. In particular, field cooled (FC) experimental curves are expected to better approach the equilibrium state of the system. Zero field cooled - Field cooled (ZFC-FC) experiments were thus conducted at several applied fields between $0$ and $1250$~Oe for the two substituted samples, as shown in Figure~\ref{ZFC_111} for the 10~\% sample. In low field (typically 100 Oe), both ZFC and FC curves show a maximum of the magnetization at a value labeled $T_{\rm m}$, which occurs roughly at the N\'eel temperature $T_{\rm N}$ (determined from the AC susceptibility). However, a splitting between the ZFC and the FC curves is observed at lower temperature, below $T_1$, with $T_1 < T_m$. As previously observed in the case of \ndhf\ \cite{Opherden_2018}, upon increasing the field, this ZFC - FC irreversibility moves to lower temperature, so that the ZFC-FC curves remain on top of each other in the temperature range $[T_1, T_m]$.

The obtained characteristic temperatures $T_1$ and $T_m$ measured at a given applied field are plotted together with the maxima $H_1$ of $dM/dH_i$ vs $H_i$ measured at constant temperature in the $(H,T)$ phase diagram shown in Figure \ref{fig_HT_111}. Interestingly, the boundary defined by the $T_1$ points is found to coincide with the $H_1$ one. Following the diffraction results of Figure \ref{res111}, the $(T_1, H_1)$ line thus separates the field induced ``1 in -- 3 out'' (or ``1 out - 3 in") regime (gray shaded region in Figure \ref{ZFC_111}) and the ``all in / all out" low field - low temperature state (blue shaded region in Figure \ref{ZFC_111}). In the latter, all magnetic moments are oriented in the same way (``in" or ``out") along the local ${\bf z}$ directions of a tetrahedron, even if the moment magnitudes are different. Since the FC curve is expected to be an equilibrium curve, $T_1$ and $H_1$ would thus correspond to the critical line where the system leaves a metastable state to reach the energetic ground state of the system upon increasing temperature or field. In that picture, the temperature $T_{\rm m}$ would correspond to the crossover between the field polarized state and the paramagnetic state. This description, which relies on the knowledge of the magnetic structure thanks to neutron diffraction, differs from the analysis of Ref.~\onlinecite{Opherden_2018} where the gray region was proposed to remain in an ``all in / all out" state, but with only one domain type. We shall come back on this point in the discussion. 

%%%%%%%%%%%%%%%%%%%%%%%%%
%DISCUSSION 
%%%%%%%%%%%%%%%%%%%%%%%%%
\section{Discussion}
\label{discussion}
\subsection{Analysis of the phase diagrams}

The phase diagrams have been calculated in Ref.~\onlinecite{Xu_2019} by Monte Carlo (MC) simulations applied to the XYZ Hamiltonian and using the coupling parameters $(\tilde{{\sf J}}_{\tilde x}, \tilde{{\sf J}}_{\tilde y}, \tilde{{\sf J}}_{\tilde z})$ available for the pure compound (See Figure~9 of this reference). These calculations were performed applying the field at high temperature and decreasing the temperature, thus corresponding to a FC procedure. Importantly, the so-called ordered parameter in these phase diagrams is associated to the AIAO component relative to the ${\bf {\tilde z}}$ directions (defined as $m_{\rm {\tilde z}\ AIAO} = \frac{1}{N}\sum_i {\tilde \tau}_i^{\tilde z}$) and not to the ${\bf z}$ components that are actually measured. 

\begin{figure}[]
\includegraphics[width=8.5cm]{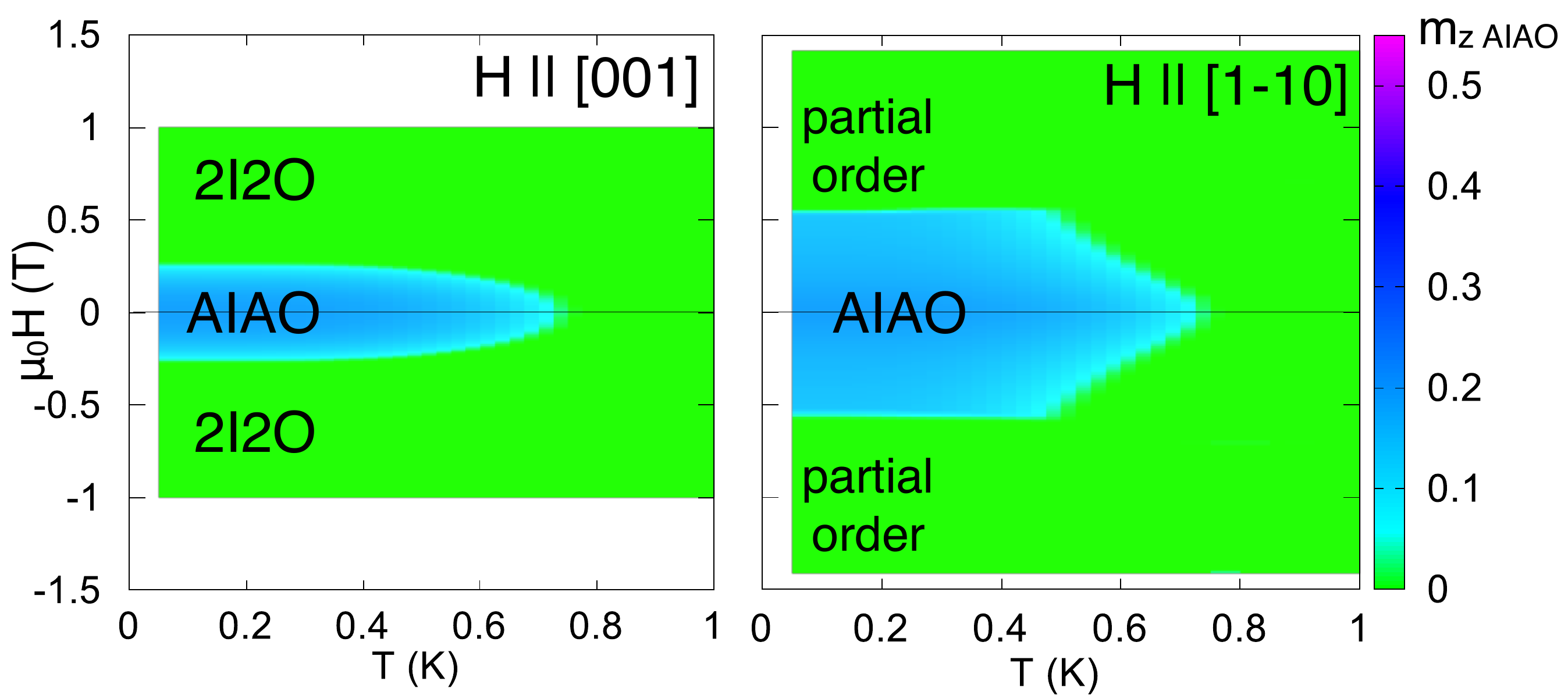}
\caption{\label{MF_001-110} Mean-field $(H, T)$ phase diagrams calculated for (a) ${\bf H} \parallel [001]$ and (b) ${\bf H} \parallel [1 \bar{1} 0]$ for the pure compound parameters. The color scale indicates the value of $m_{\rm z\ AIAO}$. Phase diagrams were obtained when sweeping the field from -1.5 T to 1.5 T.}
\end{figure}

We have performed mean-field (MF) calculations with the exchange parameters obtained for the pure compound \cite{Leger20}: $\tilde{{\sf J}}_{\tilde x}=1.18,\ \tilde{{\sf J}}_{\tilde y}=-0.03,\ \tilde{{\sf J}}_{\tilde z}=-0.53 {\rm \ K},\ \theta=1.23$ rad (See Equation \ref{ham}). Those calculations show that for the $[001]$ and $[1 {\bar 1}0]$ field directions, the phase diagrams are identical whether focusing on the ${\bf {\tilde z}}$ or ${\bf z}$-AIAO component ($m_{\rm z\ AIAO} = \frac{1}{N} \sum_i \tau_i^z$). This allows one to compare directly the calculated and measured phase diagrams. As previously obtained \cite{Xu_2018, Xu_2019}, the calculated phase diagrams qualitatively reproduce the observations but the critical fields are smaller in the experiments (see Figures \ref{resHT} and \ref{MF_001-110}). In addition, the shape of the transition line for the $[1{\bar 1}0]$ direction is quite different from the measurements. The zero field transition temperature is overestimated, as expected for mean-field calculations. 

To clarify the differences between calculations and experiments, we have followed the ``energy landscape'' approach of Ref. \onlinecite{Xu_2019}. This consists in finding the configurations which minimize the classical energy, calculated numerically using the XYZ Hamiltonian (see Appendix~\ref{landscape}). Several solutions can be found, revealing local minima and thus metastable configurations. The corresponding moments are then plotted against the magnetic field, while a color scale is used to keep track of the value of the energy. 

\begin{figure}[]
\includegraphics[width=8.5cm]{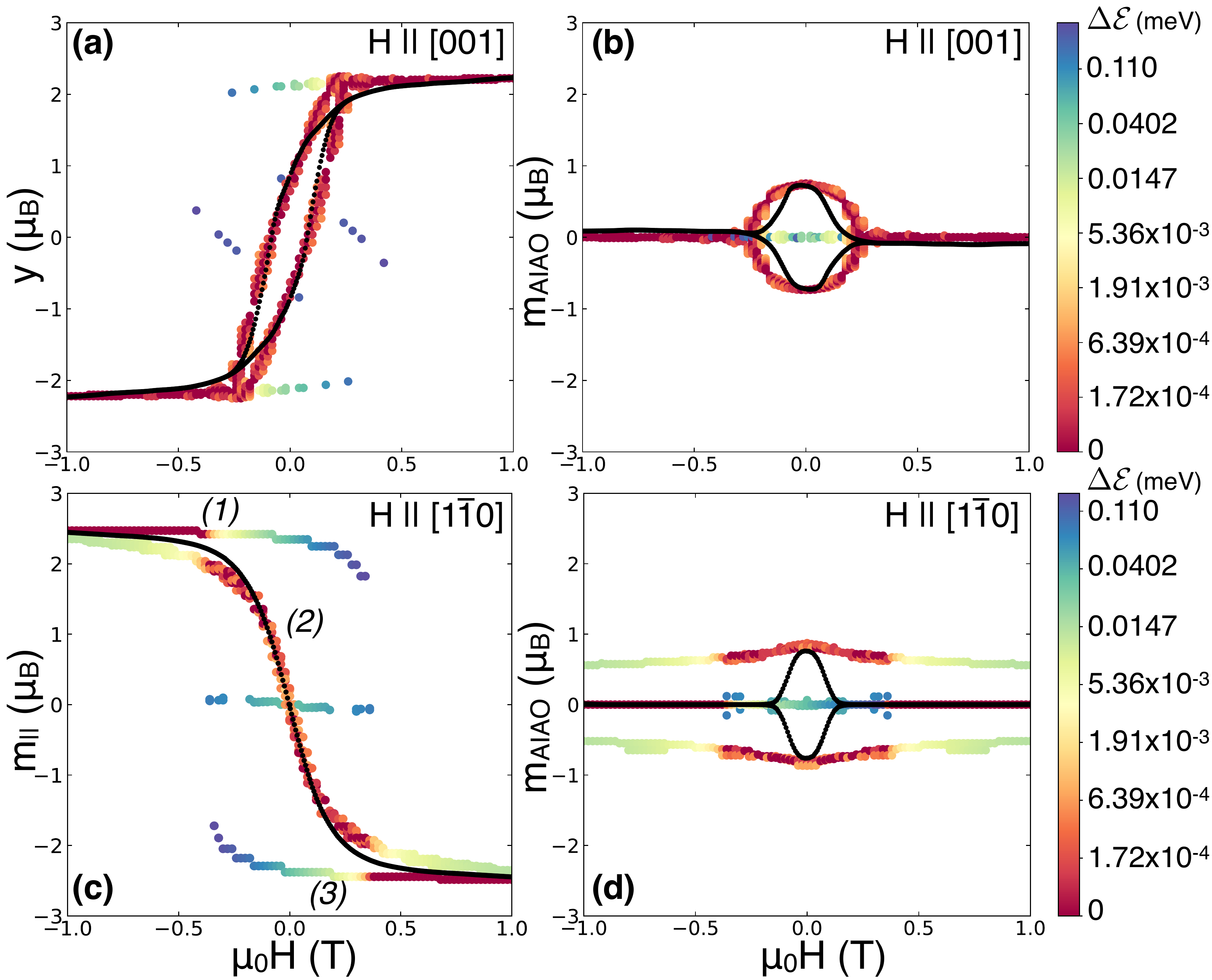}
\caption{\label{E_001-110} Magnetic components along {\bf z} as a function of field in the pure sample. (a-b) ${\bf H} \parallel [001]$: for (a) $y=m_3$ and (b) $m_{\rm AIAO}$ (see Equation \ref{m_001}). (c-d) ${\bf H} \parallel [1 {\bar 1} 0]$: for (c) $m_{\parallel}$ and (b) $m_{\rm AIAO}$ (see Equation \ref{m_110}). The color scale represents the energy difference between the calculated configuration and the absolute minimum energy configuration (logarithmic scale), the red color refering to the smallest energy. Black dots correspond to the experimental values obtained from the field sweeping measurements analysis. Labels (1,2,3) on panel (d) refers to the three branches discussed in the text. }
\end{figure}

Figure \ref{E_001-110}(a-b) shows the results for the field applied along the $[001]$ direction together with the diffraction results in the pure sample. As expected from a simple symmetry analysis and the neutron results (discussed in Section \ref{001}), two magnetic configurations are equivalent in energy and define two branches for the evolution of magnetic moments {\it vs} field. This can be seen by plotting the magnetic component $y$ of the Nd3 moment (panel~(a)), or the $m_{\rm z\ AIAO}$ component (panel (b)). Two domains then probably form in the system when ramping the field, that cannot be distinguished in experiments. The field value at which the two branches merge, which corresponds to the suppression of the AIAO component, is slightly larger in the calculation than in the experiment, but roughly in the same range. The slight hysteresis observed experimentally may be due to a small field misalignment which would favor one domain over the other, resulting in metastable configurations, and reducing the critical field, in a similar way as for the $[111]$ direction that we will discuss below. 

In the $[1\bar{1}0]$ direction, the agreement between the calculations and the experiments is less convincing as can be seen in the panels (c-d) of Figure \ref{E_001-110}. The experimental $m_{\rm z\ AIAO}$ (and thus the two domains) vanishes at much smaller field and more smoothly than in the calculations (panel (d)) where an abrupt jump is obtained.  When focusing on the $\alpha$ chains (Nd3 and Nd4), calculations for the parallel component $m_{\parallel}$ (panel d) predict essentially three branches, labeled (1,2,3) on the figure: branches (1) and (2) correspond to an almost saturated $m_{\parallel}$ moment, and are stable at high values of the (absolute) field. The central branch (3) is stable between about $-0.3$ and 0.3~T. Looking at the color scale, and starting from $-1$~T, the global energy minimum thus goes from a saturation branch to the other via the central branch. In this part,  $m_{\rm z\ AIAO} \neq 0$ and the moment decreases smoothly. While in the calculations a little step is obtained when going from one branch to the other, concomitantly with the appearance and suppression of the ``all in" or ``all out" component, the experimental data seem to go continuously from one branch to the other. 

A small jump is nevertheless observed in the derivative of the magnetization, but it is hysteretic and much weaker than in the present calculations. The discrepancy between the calculations and the experiments may lie in the subtlety of the $[1\bar{1}0]$ situation: the two $\beta$ spins which remain disordered in the nearest neighbor classical approach may be sensitive to octupolar and long range dipolar couplings, or experience order by disorder effects \cite{Xu_2018, Guruciaga16, Placke20}. It is clear from our results that these ingredients are not strong enough to affect the high field state since the Nd1 and Nd2 moments do not eventually order (which also excludes a strong misalignment of the field), but they may affect the low field picture. \\

\begin{figure}[!t]
\includegraphics[width=8.5cm]{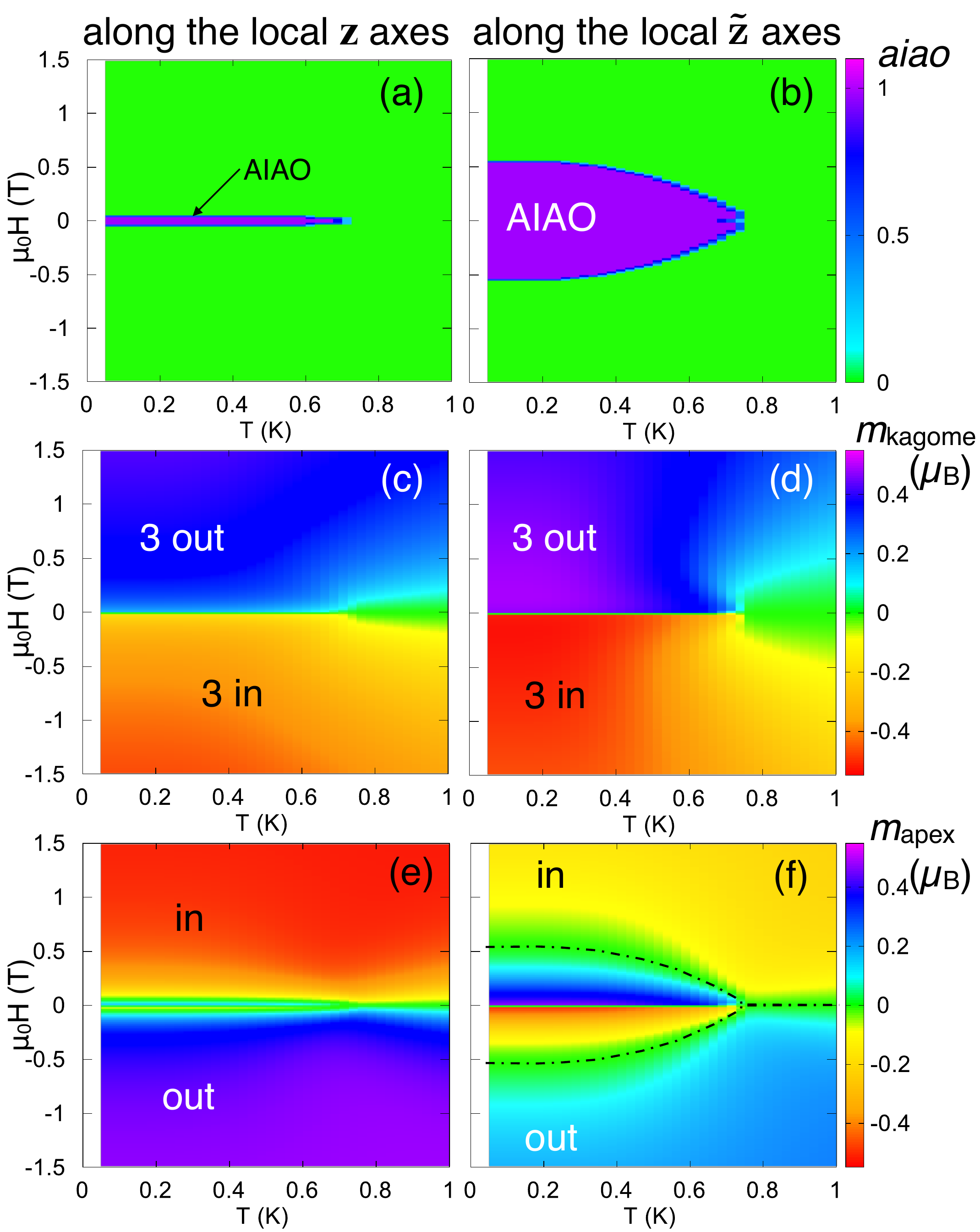}
\caption{\label{MF_111_FC} Mean-field $(H,T)$ phase diagrams obtained for ${\bf H} \parallel [111]$ upon decreasing the temperature in field cooled conditions for the pure compound parameters. The left and right sides show the same results but projected along the $\bf z$ and $\bf \tilde z$ components of the pseudo spins respectively. (a-b) show the $aiao$ parameter (equal to 1 if all the spins are ``in" (or ``out") and 0 otherwise). (c-d) display the kagom\'e spins component $m_{\rm kagome}$, (e-f) the apical spin component $m_{\rm apex}$. Dashed lines are guides to the eye to highlight the region delimited by the $aiao$ parameter in (b). }
\end{figure}

\begin{figure}[!t]
\includegraphics[width=8.5cm]{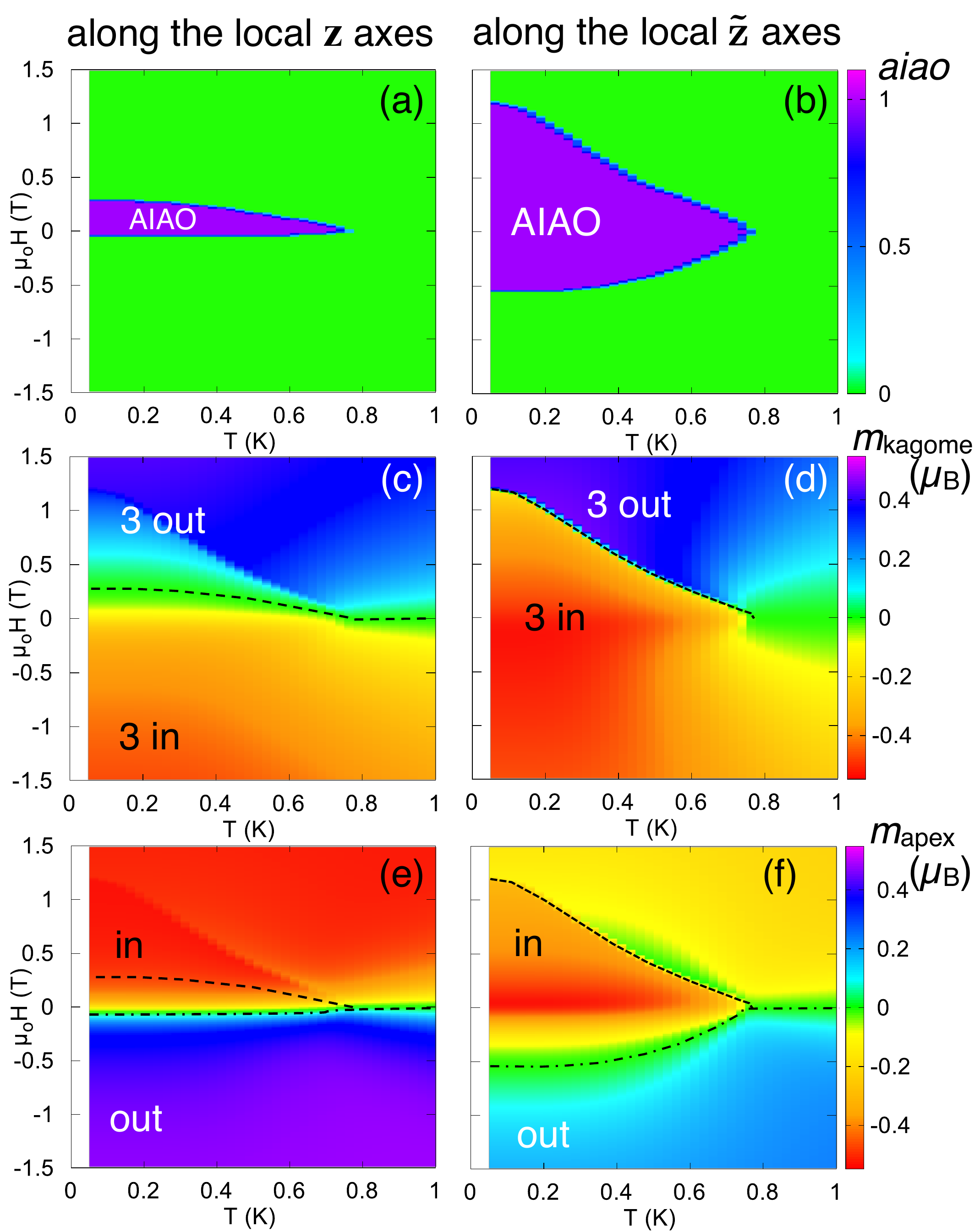}
\caption{\label{MF_111_hyst} Mean-field $(H,T)$ phase diagrams obtained for ${\bf H} \parallel [111]$ when the field is swept from $-1.5$ to $1.5$ T for the pure compound parameters. The left and right sides show the same results but projected along the $\bf z$ and $\bf \tilde z$ components of the pseudo spins respectively. (a-b) displays the $aiao$ parameter, (c-d) the kagom\'e spins component $m_{\rm kagome}$, (e-f) the apical spin component $m_{\rm apex}$. Dashed lines are guides to the eye to show the regions delimited by the $aiao$ parameter in (a) and (b).}
\end{figure}
We now turn to the $[111]$ direction, where the description of the field induced process is less direct. Indeed, as already mentioned, the description in terms of an AIAO state is not relevant, due to the different fields felt by the apical and kagom\'e spins, the main consequence being that all the moments do not have the same ``length'' along the field direction. To get a clear insight of what happens in this case, it is convenient to define the $aiao$ parameter, which is equal to 1 if all the spins are ``in" (or ``out") and 0 otherwise. This parameter can have a different value depending on whether it is defined with respect to the ${\bf \tilde z}$ or to the ${\bf z}$ directions, as illustrated in Figures~\ref{MF_111_FC} and \ref{MF_111_hyst} (a-b). Actually this is of great importance to understand the measured magnetic structure and phase diagrams. 

At first, it appears simpler to discuss the phase diagram measured and calculated in field cooled conditions, since it is expected to describe the equilibrium state, and get rid of metastable states. Mean-field calculations shown in Figure \ref{MF_111_FC} highlight two descriptions. The left column represents the $aiao$ parameter and magnetic components along ${\bf z}$, which can be directly related to the experiments. The right column shows the $aiao$ parameter and pseudo spin components along ${\bf \tilde z}$. Those are not observed in experiments, but we know from calculations that they are fully AIAO ordered in zero field. In the following, a distinction will be made between the configurations with respect to ${\bf z}$ (${\bf z}$-AIAO for instance) and to ${\bf \tilde z}$ (denoted ${\bf \tilde z}$-AIAO).
 
\begin{figure}[h!]
\includegraphics[width=8cm]{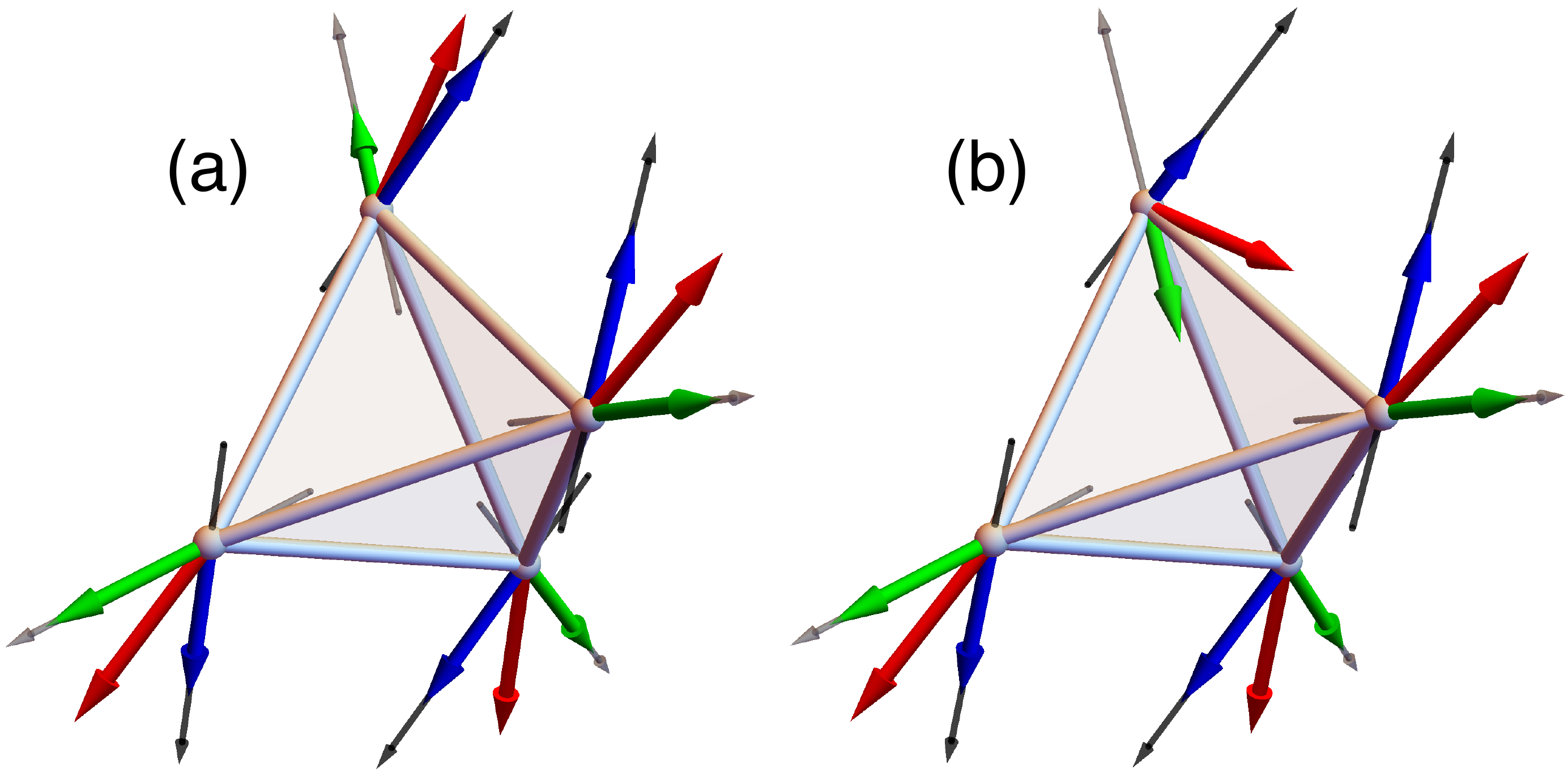}
\caption{\label{mag_struct} Sketch of the pseudo spins in a tetrahedron (a) in the blue region and (b) in the gray region of the phase diagram. The black and gray arrows respectively represent the ${\bf \tilde z}$ and ${\bf z}$ local axes. Red arrows represent the total pseudo spin components, the blue ones their projection on the ${\bf \tilde z}$ axes and the green ones their projection on the ${\bf z}$ axes. In (a), an ``all out" configuration is obtained for both projections, while in (b), only the ${\bf \tilde z}$ components are ``all out", the ${\bf z}$ ones being 1O3I.}
\end{figure}

Calculations show that the magnetic state is governed by the apical spins, as the kagom\'e spins simply follow the field direction (panels (c) and (d) of Figure \ref{MF_111_FC}). On the contrary, the apical pseudo spin is essentially along ${\bf \tilde{z}}$ at low field, typically below $0.5$ T, hence forming tetrahedra either  ${\bf \tilde{z}}$-``all in'' or  ${\bf \tilde{z}}$ -``all out''. Above this threshold, the tetrahedra are ${\bf \tilde{z}}$-1I3O or ${\bf \tilde{z}}$-1O3I depending on the sign of the field. Interestingly, however, even if the pseudo spins are globally ${\bf \tilde{z}}$-AIAO, the actual magnetic configuration, i.e. with respect to ${\bf z}$, can be ${\bf z}$-1I3O (or ${\bf z}$-1O3I) (see Figure \ref{mag_struct}). When comparing quantitatively with the experimental data obtained from field cooled measurements (see Figure \ref{fig_HT_111}), the entrance in the gray region delimited by $T_{\rm m}$ (maximum of the susceptibility) upon cooling seems to correspond to the entrance into a ${\bf \tilde{z}}$-AIAO state.

\begin{figure}[!h]
\includegraphics[width=8.5cm]{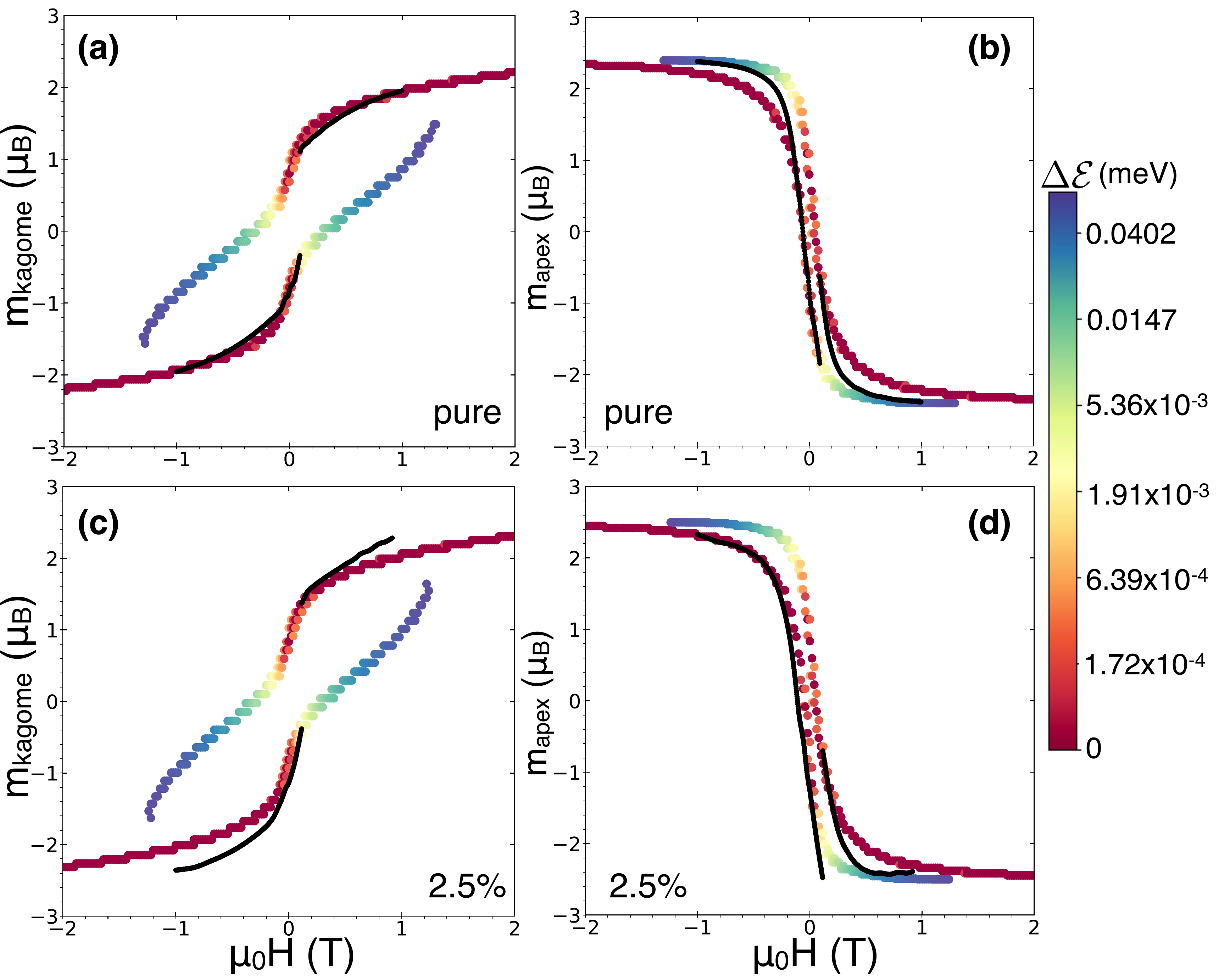}
\caption{\label{E_111} Magnetic moment as a function of field for the pure (a-b) and the 2.5 \%(c-d) compounds when a field is applied along $[111]$ for the (a-c) kagom\'e and (b-d) the apical spins. The color scale represents the energy difference between the calculated configuration and the absolute minimum energy configuration (logarithmic scale), the red color refering to the smallest energy. Black dots correspond to the experimental values obtained from the field sweeping measurements analysis. }
\end{figure}

Similar calculations were performed by starting from a saturated state in negative field and increasing the field, similarly to the hysteresis loop / field sweeping experiments (see Figure \ref{MF_111_hyst}). These calculations clearly reproduce the hysteretic behavior observed experimentally (see Figure \ref{res111}). A transition is observed at $1.2$ T (right panels). It corresponds to the flipping of the kagom\'e spins from a ${\bf \tilde z}$-``in" to a ${\bf \tilde z}$-``out" configuration, and to a reduction of the ordered value for the apical spin. The change from the ${\bf z}$-AIAO to the ${\bf z}$-1I3O state, however, occurs gradually, the ${\bf z}$-AIAO component disappearing around $0.25$ T (left panels). These calculations thus cannot reproduce the transition measured at about $0.1$ T. To understand this discrepancy, which was not observed for the field cooled case, the method of the energy landscapes was used again (see Appendix \ref{landscape}). As inferred from the experimental results, and previously discussed \cite{Xu_2019}, this approach reveals the presence of two local energy minima, which result in metastable states when sweeping the field. Figure \ref{E_111}(a-b) shows the magnetic moment value associated to these local minima for the pure compound, while the color scale indicates the energy of the associated configuration. When comparing with mean-field calculations presented in Figure \ref{MF_111_hyst}, these results reveal that the magnetic moment remain ``trapped'' on a single branch, which becomes metastable for positive fields. As previously noted in Ref. \onlinecite{Xu_2019}, the jump to the lowest energy branch, i.e. the critical field obtained in the mean-field approach corresponds to the point ($\mu_0H=1.2$~T) where the metastable branch disappears. 
The experimental magnetic moments obtained for the pure compound from the neutron diffraction data are displayed on top of these calculated points in Figure \ref{E_111}(a-b). Remarkably, the experimental points, especially for the kagom\'e moment, nearly fall on the calculated branches, suggesting that everything happens as if the system was jumping from the metastable branch to the minimum energy branch at a much smaller field ($H_{111}$), following from there the absolute energy minimum. The magnetic configuration at this jump (for the red tetrahedra) changes from a ${\bf z}$-``all in" state to a ${\bf z}$-1I3O configuration. The system thus reaches its equilibrium state at a much smaller field than predicted in MF and MC calculations. The associated gain in energy obtained by jumping from one branch to another at $H_{111}$ is found to be about 50 mK. This shows the existence of relaxation processes not captured by calculations. 

Experimentally, we have pointed out that $H_1$ (determined from $M(H)$ curves) and $T_1$ (determined from ZFC-FC curves) define the same line in the $(H,T)$ phase diagram. Indeed, in a ZFC measurement, both ${\rm \overline{AIAO}} / {\rm \overline{AOAI}}$ domains are stabilized at low temperature. When applying a positive field (with our conventions) after ZFC, the ``all out" domains of the red tetrahedra are on the equilibrium branch while the ``all in" ones are on the metastable branch.  When increasing the field and/or the temperature, the latter are suppressed. The system then reaches the equilibrium configuration, when the ZFC magnetization recovers the FC equilibrium one. This occurs at the same field / temperature point in the experimental phase diagram as the magnetization jump in the hysteresis. Incidentally, at low temperature, it roughly corresponds to the point where the ${\bf z}$-AIAO component disappears in the calculated field cooled phase diagram.

Finally, we can now understand the nature of the experimental phase diagram of Figure \ref{fig_HT_111}, when the field is applied along $[111]$ (the measurements being performed from a ZFC state, or a saturated state in a negative field): in the blue region both ${\bf z}$ and ${\bf \tilde z}$ components are AIAO (see the illustration in Figure \ref{mag_struct}(a)), and the system lies in a metastable state. In the gray one, only the ${\bf \tilde z}$ component is AIAO, the ${\bf z}$ one being 1I3O / 1O3I (see the illustration in Figure \ref{mag_struct}(b)), and the system has reached its ground state. The microscopic mechanism at the origin of the measured values of $H_1$ and $T_1$ separating these two regions nevertheless remains an issue.

\subsection{Role of the Ti content}

In this study, the $(H,T)$ phase diagrams have been established for three samples, the pure \ndzr\ compound and two Ti-substituted samples, Nd$_2($Zr$_{1-x}$Ti$_x)_2$O$_7$ with $x=2.5$ and 10 \%. As shown in Table \ref{Table_charac}, the global pyrochlore stucture and its main properties (lattice parameter and oxygen position) are not  significantly affected by the Ti substitution at low concentrations of Ti. Further structural studies are necessary to better characterize the disorder, but are beyond the scope of this article. From the present data, however, it is reasonable to consider that the titanium atoms are randomly distributed among the 16d sites of the crystallographic Fd${\bar{3}}$m structure. 

It is worth noting that the properties of the substituted polycrystalline samples are quantitatively different from the single crystal ones (smaller $\theta_{\rm CW}$ and slightly larger $T_{\rm N}$ and ordered moments - see Appendix \ref{powder}). In the literature, larger $T_{\rm N}$ and ordered moments were also reported in a pure single crystal compound \cite{Xu_2015}. The nature of the disorder and its consequences are thus not obvious, but nevertheless, two important features can be stressed. (i) The single ion properties of the neodymium ion are little affected by the presence of this disorder, the effective moment and local anisotropy being roughly similar in the three compounds. (ii) The presence of titanium is expected to mainly disturb the exchange paths, creating some distribution in the exchange parameters.

Indeed, the magnetic transition looks a bit broader in the AC susceptibility (see Figure \ref{Xac}) for substituted samples, but is still well defined. Strikingly, the AIAO ground state is characterized by a larger $T_{\rm N}$ along with larger critical fields than in the pure sample, suggesting that the AIAO order is more robust. Since both characteristic parameters show a maximum at 2.5 \% before decreasing at higher substitution concentration, the present study even suggests the existence of an optimum with respect to Ti-substitution level, where the stability of the ordered phase is best reinforced. In addition, the ordered magnetic moment is larger along the ${\bf z}$ direction, thus corresponding to a lower $\theta$ angle, which indicates that the disorder tends to tilt the pseudo spins back towards the local $\langle 111 \rangle$ ${\bf z}$ axes. 
This picture is consistent with recent observations on Nd-based pyrochlores with a much stronger disorder on the $B$-site \cite{Mauws19, Scheie21, Gomez21}, namely  Nd$_2$NbScO$_7$ and Nd$_2$GaSbO$_7$. In the latter, it was nevertheless pointed out that the large obtained ordered moment and transition temperature are similar to the Nd$_2$Sn$_2$O$_7$ compound \cite{Bertin15}. It was thus proposed that the chemical pressure - which manifests in these two compounds by a small lattice parameter, compared to the zirconate counterpart - is a key ingredient to understand the magnetic properties, and which could be more decisive than the disorder itself. 

\begin{figure}[h!]
\includegraphics[width=8.5cm]{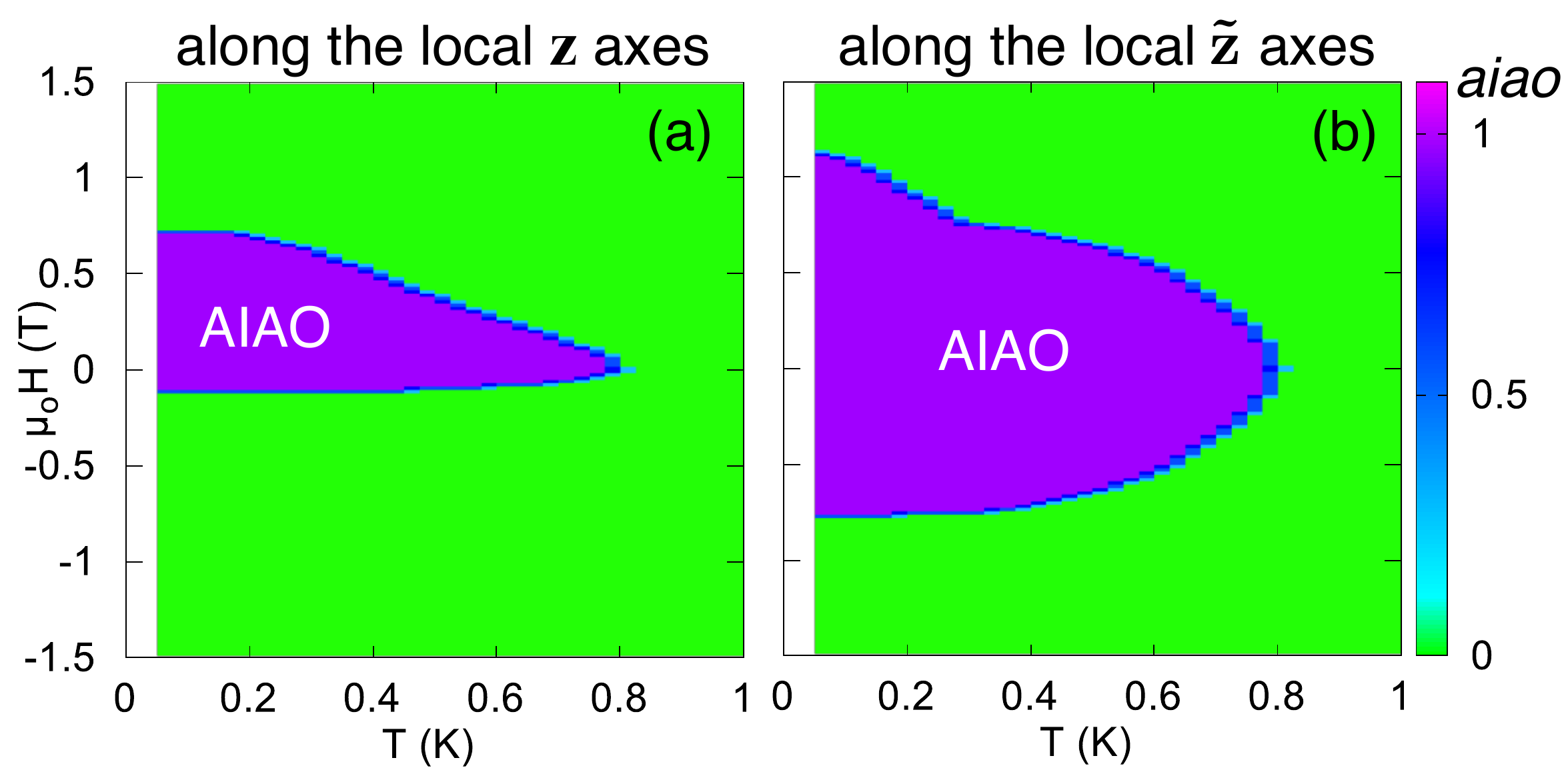}
\caption{\label{calc_doped}
(a,b)  $aiao$ parameter vs $(H,T)$ determined from mean-field calculations considering ${\bf H} \parallel [111]$ and swept from $-1.5$ to $1.5$ T performed with the exchange parameters determined for the 2.5 \% substituted sample. (a) (respectively b) shows the results for the $\bf z$ (respectively $\bf \tilde z$) component of the pseudo spins.}
\end{figure}

Despite the absence of a microscopic description of the disorder which could help solving this issue, the phase diagram of the 2.5 \% substituted sample can be further analyzed. To this end, we have computed the MF phase diagrams as well as the energy landscape using the exchange parameters determined by the magnetic excitations analysis \cite{Leger20} (see Figure \ref{calc_doped}(a,b)). The obtained values are $\tilde{{\sf J}}_{\tilde x}=0.97,\ \tilde{{\sf J}}_{\tilde y}=0.21,\ \tilde{{\sf J}}_{\tilde z}=-0.53 {\rm \ K},\ \theta=1.08$ rad (See Equation \ref{ham}. For the sake of simplicity, $\tilde{{\sf J}}_{\tilde y}=0$ was assumed for energy landscape calculations). 

The first point that we stress is that, in mean-field calculations, the N\'eel temperature does not vary significantly between the results for the pure and 2.5 \% substituted compounds. This is not surprising since the energy range of the parameters is similar in both, and the calculations strongly overestimate $T_{\rm N}$ (while MC calculations underestimate it \cite{Xu_2019}). Nevertheless, experimentally, a strong variation of \TN\ is observed between the three samples. This may be due to the competition at play between the Coulomb phase stabilized above the N\'eel temperature for the ${\bf \tilde x}$ component and the ${\bf \tilde z}$-AIAO phase \cite{Xu_2019, Leger20} (see also Appendix \ref{frag}). This reminds the situation in Yb$_2$Ti$_2$O$_7$ \cite{Arpino_2017, Robert15, Jaubert15, Scheie20}, where the proximity of competing phases makes the transition temperature extremely sensitive to disorder. In \ndzr, the disorder would promote the AIAO ordering at the expense of the Coulomb phase. 
This manifests in the exchange parameters by a smaller ratio $\tilde{{\sf J}}_{\tilde x} / \tilde{{\sf J}}_{\tilde z}$ in the substituted compound which is clearly in favor of the ${\bf \tilde z}$-AIAO phase.

When looking at the field induced behavior, calculations qualitatively reproduce the increase of the critical fields obtained experimentally in the $[001]$ and $[1\bar{1}0]$ directions, as well as the nature of the field induced processes, which are similar to the pure compound. Nevertheless, there is no quantitative agreement in the obtained values for the critical field (see Table \ref{Table_HM}). When the field is applied along the $[111]$ direction, an increase of the critical field is obtained in the field cooled phase diagrams. Considering now the situation where the field is swept from saturation, almost no difference is obtained in the calculated critical field (of about 1.2 T) corresponding to the jump from a metastable domain to the equilibrium. When looking in more details at the branches displayed in Figure \ref{E_111}(c,d), we can see that their shape is quite different from the pure compound case, even if the two branches merge at almost the same field of about 1.2 T. The data obtained from the neutron field sweeping analysis falls again on the branches, even if the agreement is less good for the kagom\'e moment in high field. The jump from one branch to the other occurs experimentally at $0.11$ T, which corresponds to an energy difference of 50 mK, comparable to the case of the pure compound. 

From our experimental and calculation results, the samples comparison thus shows that despite quite strong quantitative differences on the N\'eel temperature or on the ${\bf z}$ ordered moment, the field induced behaviors remain similar, the critical fields being shifted to larger values in the substituted samples. The level of disorder introduced in the present study on the non magnetic site is not strong enough to destabilize the ${\bf \tilde z}$-AIAO ordering, and on the contrary seems to reinforce it. As previously mentioned, recent studies show that the AIAO state resists to an even stronger disorder \cite{Mauws19, Scheie21, Gomez21}. Consequently, it may be relevant to directly make a substitution on the magnetic site in order to test the stability of the AIAO state on dilution, and to possibly reach the predicted quantum spin liquid ground state when the ratio $\tilde J_{\tilde z}/ \tilde J_{\tilde x}$ decreases \cite{Benton_2016, Benton2020}.

%%%%%%%%%%%%%%%%%%%%%%%%%
%Conclusion
%%%%%%%%%%%%%%%%%%%%%%%%%
\section{Conclusion}
In this work, $(H,T)$ phase diagrams for fields applied along the three main high symmetry crystallographic directions have been constructed for three sample compositions of the \ndzrx\ family. A systematic analysis of neutron diffraction data was employed, combined with low temperature magnetization measurements. Depending on the field direction, different phases are obtained, either AIAO, 2I2O or 1I3O / 1O3I. The present work essentially establishes the stability regions of the AIAO phase, and shows that a small amount of disorder further reinforces this phase, both in field and temperature. In addition, the comparison with calculations performed using the XYZ Hamiltonian shows how the field induced phases (2I2O and 1I3O / 1O3I) are connected to the two ${\rm \overline{AIAO}}$ and ${\rm \overline{AOAI}}$ domains, and elucidates the apparent strong discrepancy between calculated and measured critical fields when the field is applied along the $[111]$ direction. Further work on the microscopic mechanisms at play in the domain selection are nevertheless needed to quantitatively understand the critical field values.  

%%%%%%%%%%%%%%%%%%%%%%%%%%%%%%%%%%%%%%%%
% Acknowledgment
%%%%%%%%%%%%%%%%%%%%%%%%%%%%%%%%%%%%%%%%
\acknowledgments
The work at the University of Warwick was supported by EPSRC, UK through Grant EP/T005963/1. M.L. and S.P. acknowledge financial support from the French Federation of Neutron Scattering (2FDN) and Universit\'e Grenoble-Alpes (UGA). F.D, E.L. and S.P. acknowledge financial support from ANR, France, Grant No. ANR-19-CE30-0040-02. 

%%%%%%%%%%%%%%%%%%%%%%%%%%%%%%%%%%%%%%%%
%Appendix
%%%%%%%%%%%%%%%%%%%%%%%%%%%%%%%%%%%%%%%%
\appendix
\begin{table*}[t]
\begin{tabularx}{\textwidth}{cYcY*{4}{c}}
\hline
\hline
\multirow{2}{*}{Powder Samples} & Ti concentration & \multirow{2}{*}{O2 position} & Lattice parameter & $T_{\rm N}$ & $\theta _{\rm CW}$ & $\mu _{\rm eff}$ & Ordered moment \\
& [\%] && [\AA] & [mK] & [mK] & [$\mu _{\rm B}$] & [$\mu _{\rm B}$] \\
\hline
%1
$\mathrm{Nd_2Zr_2O_7}$                       & 0                      &0.3366 & 10.69 & 370 & 225 & 2.42 & $0.8 \pm 0.05$ \\
%2
$\mathrm{Nd_2Zr_{1.95}Ti_{0.05}O_7}$ & 2.8  &0.3360  & 10.67 & 420 & 70 & 2.53 &$1.36 \pm 0.04$\\ % \pm 20
%3
$\mathrm{Nd_2Zr_{1.8}Ti_{0.2}O_7}$   & 11.2  & 0.3359 & 10.64 & 430 & 80 & 2.51 & $1.33 \pm 0.03$ \\ %\pm 20
\hline
\hline
\end{tabularx}
\caption{\label{table_powder} Polycrystalline parameters obtained from neutron diffraction and magnetization measurements.  Structural parameters are from {\sc Fullprof} refinements at 300 K and the ordered moment is the value obtained at about 60 mK. }
\end{table*}

\section{Notes about fragmentation}
\label{frag}
Prior experimental investigations of \ndzr\ \cite{Lhotel_2015,Petit_2016}, have pointed to the fact that the ground state of this material bears similarities with the fragmentation phenomenon exposed in Ref. \onlinecite{Brooks_2014}. In particular, the fact that only 1/3 of the total moment is ordered was one of the strongest arguments.\\

In this theory, the proliferation of spin flips out of the spin ice 2I2O manifold, i.e. 1O3I and 1I3O local configurations, or monopoles in the field language, produces a ``charged'' state, described by two fragments that identify with the Helmholtz-Hodge decomposition of the spin ice emergent magnetic field ${\bf M}$: 
\begin{equation*}
{\bf M} = - {\bf \nabla} \phi + {\bf \nabla} \times {\bf A} 
\end{equation*}
where $\phi$ is a scalar potential, and ${\bf A}$ the vector potential. In this decomposition, the first term is divergence-full and carries the total gauge charge $\rho$, ${\bf \nabla} \cdot {\bf M} = -{\bf \nabla} \cdot {\bf \nabla} \phi = \rho$. The second term $\nabla \times {\bf A}$ encodes the position of the spin flip and is divergence-less (${\bf \nabla} \cdot ({\bf \nabla} \times {\bf A}) \equiv 0$). In the case of spin ice, the ground state is naturally the vacuum of charges ($\rho=0$). However, a different ground state appears if a non-zero density of monopoles becomes favored for energetic and/or entropic reasons. Provided the fluid of monopoles crystallizes and forms a staggered pattern, the new ground state is fragmented, comprised of an AIAO component carried by the divergence full term on top of a secondary Coulomb phase carried by the divergence free term. Importantly, the charge per tetrahedron ($\rho=2$) is only half the charge carried by a standard AIAO state ($\rho=4$). As a result, in a neutron diffraction experiment, the fragmented state is characterized by an AIAO ordering with an ordered magnetic moment reduced by a factor of two, on top of a spin ice pattern described by ${\bf A}$. 

The XYZ Hamiltonian approach is quite far away from this theory of fragmentation. Its phase diagram was studied using the gauge mean field theory (gMFT) in Ref. \onlinecite{Huang14,Li17}. It presents unconventional $U(1)$ quantum spin liquid states \cite{Hermele04,Shannon12,Benton12,Savary12,Savary13,Hao14,Huang18}, but also includes classical states. These are ordered AIAO-like phases, with pseudo spins pointing along the $a=\tilde{x},\tilde{y},\tilde{z}$ direction. At the mean field level, they are stabilized provided $\tilde{{\sf J}}_{a} < 0$ with $-1 \le \frac{\tilde{{\sf J}}_{b,c}}{|\tilde{{\sf J}}_{a}|} \le 3$. In Ref. \onlinecite{Huang14}, those states are labeled AIAO and AFO (the latter corresponding to an antiferro-octupolar state). Interestingly, the experimental couplings place \ndzr\ close to the border between the $\tilde{x}-U(1)$ liquid phase and the $\tilde{z}$-AIAO state in the gMFT phase diagram determined by Huang {\it et al.} \cite{Huang14}. The most recent experiments actually support a picture where the paramagnetic regime above the AIAO N\'eel temperature in \ndzr\ is indeed such a coulombic phase \cite{Xu_2020, Leger20}, stabilized both by the strong positive $\tilde{{\sf J}}_{x}$ as well as entropic effects. 

Ref. \onlinecite{Benton_2016} describes a mean field theory of this XYZ model. For the \ndzr\ parameters, the ground state is an AIAO magnetic configuration of the pseudo spins pointing along the $\tilde{z}$ axes and is thus a fully charged state. However, it turns out that the {\it dynamical} magnetization can still be ``Helmholtz-Hodge'' decomposed in terms of (lattice) divergence-free and divergence-full fields \cite{Benton_2016}. On the one hand, the flat spin ice mode identifies with the divergence-free {\it dynamical} field. Its pinch point pattern relates it to the divergence free condition ${\bf \nabla} \cdot ({\bf \nabla} \times {\bf A}) \equiv 0$. It is formed by individual precessions around the local magnetization of spins belonging to closed loops in the pyrochlore lattice. On the other hand, the dispersing branch corresponds to the divergence-full {\it dynamical} field and is the signature of charge propagation throughout the diamond lattice. This branch is characterized by a beautiful half moon pattern when moving away from the pinch points. Such a pattern is typical of a curl free condition, which characterizes the charges, since ${\bf \nabla} \times ({\bf \nabla} \phi) \equiv 0$ \cite{Yan_2018}. 

%%%%%%%%%%%%%%%%%%%
\section{Powder samples}
\label{powder}

Aiming at a first description of the physical properties of the substituted samples, magnetic measurements along with neutron diffraction were performed on the substituted powder samples. The very low temperature magnetization and susceptibility were measured in the same magnetometer developed at the Institut N\'eel as single crystals. For these measurements, the powder samples were packed in a copper pouch with apiezon grease to ensure proper thermalization. Structural parameters were obtained with the high resolution powder neutron D2B (ILL) diffractometer using $\lambda = 1.594$~\AA. The ground state magnetic structure in zero applied field was determined on the powder neutron G4.1 (LLB-Orph\'ee, France) diffractometer using $\lambda = 2.43$ \AA. 
%Powder neutron diffraction experiments were carried out at G4.1 (LLB-Orph\'ee, France) using $\lambda = 2.43$ \AA\ to determine the ground state magnetic structure in zero applied field. 
A dedicated vanadium sample holder was used, loaded with $^4$He pressure of about $10$ bars to ensure thermalization. Structural and magnetic parameters were refined with the {\sc Fullprof} suite \cite{Fullprof}.

\begin{figure}[h!]
\includegraphics[width=8cm]{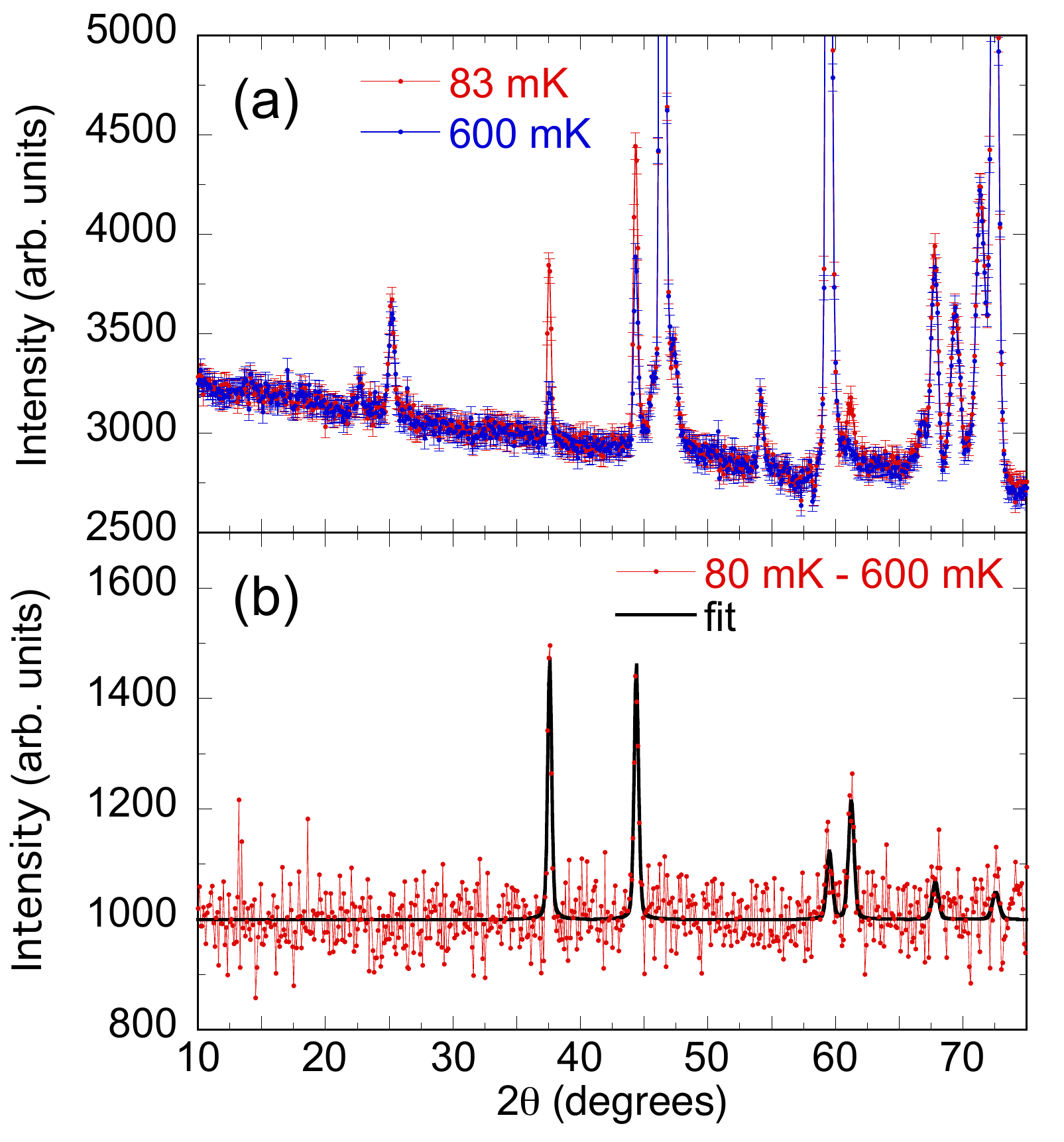}
\caption{\label{magstruct_powder} (a) Powder neutron diffractograms measured on G4.1 for the $x=2.5$ \% sample at 80 mK (red dots) and 600 mK (blue dots). (b) Difference intensity obtained when subtracting the 600 mK to the 80 mK data (red dots). The black line shows the AIAO {\sc Fullprof} refinement, giving an ordered moment of 1.36 $\mu_{\rm B}$. }
\end{figure}

The structural data could be refined with a pyrochlore structure (Fd$\bar{3}$m space group). Magnetic properties are qualitatively similar to the pure compound properties (see Table \ref{table_powder}): from Curie-Weiss fits between $1.5$ and $4$ K, the low temperature effective moment is estimated to $2.5~\mu_{\rm B}$ and a small (70 - 80 mK) but positive Curie-Weiss temperature could be extracted, indicative of effective ferromagnetic interactions. An antiferromagnetic transition is found to occur above 400 mK for both compounds in the susceptibility, but which is sharper in the 2.5 \%-substituted sample. Neutron diffraction refinements could show that the magnetic structure is AIAO, like in the pure compound, with an ordered moment of about $1.2-1.3~\mu_{\rm B}$ at low temperature (See Figure \ref{magstruct_powder}). 

%%%%%%%%%%%%%%%%%%%
\section{Analysis of the field sweeping measurements}
\label{ramp_analysis}
\begin{table}[h!]
\renewcommand{\arraystretch}{1.2}
{\begin{tabular}{ccP{6cm}}
\hline
\hline
$x$ & ${\bf H}$ & ${\bf Q}$ \\
\hline
\hline
\multirow{4}{*}{0} & [001] & (2,2,0), (0,4,0), (1,1,1), (2,0,0), (3,1,-1) \\
& [1-10] & (0,0,2), (0,0,4), (1,1,3), (2,2,0), (3,3,1), (1,1,1) \\
& [111]  & (2,0,0), (2,-2,0), (1,-1,1), (1,3,-1) \\ 
\hline 
\multirow{5}{*}{2.5 \%} & [001] & (2,-2,0), (0,4,0), (-1,-1,-1), (-1,1,-1), (-3,1,-1), (-1,-3,-1), (-3,-3,-1) \\
& [1-10] & (0,0,2), (0,0,4), (1,1,3), (2,2,0), (3,3,1), (3,3,-1) \\
& [111]  & (-2,0,0), (-2,2,0), (-1,1,-1), (-1,-3,1) \\ 
\hline 
\multirow{3}{*}{10 \%} & [001] & (2,-2,0), (0,4,0), (1,-1,1), (2,0,0), (3,1,1), (-3,3,1) \\
%& [1-10] & ? \\
 & [111]  & (-2,0,0), (-2,2,0), (-1,1,-1), (-1,-3,1) \\ 
\hline
\hline
\end{tabular}}
\caption{\label{table_Q} List of the ${\bf Q}$ vectors of the magnetic Bragg peaks measured in different experiments while ramping the magnetic field.}
\end{table}

To complete the diffraction measurements, the intensity of selected Bragg peaks was recorded while ramping the field back and forth from $-1$ T to $1$ T ($0.015$~T/min was the minimum sweeping speed) to obtain a ``continuous'' evolution vs field. The analysis of those field sweeping measurements relies on the comparison between the experimental magnetic intensities $I^{\mbox{{\small exp}}}({\bf Q})$ and the calculated ones $I^{\mbox{{\small calc}}}({\bf Q})$, given the various models described in the main text and the usual definitions of the neutron intensities:
\begin{eqnarray*}
I^{\mbox{{\small calc}}}({\bf Q}) &=& \sum_{a,b=x,y,z}
{\bf F}^a_{\rm M}({\bf Q})
\left(\delta_{a,b}-\frac{{\bf Q}^a {\bf Q}^b}{{\bf Q}^2}\right)
{\bf F}^b_{\rm M}({\bf Q})^* \\
{\bf F}_{\rm M}({\bf Q}) &=& \sum_i {\bf m}_i~e^{i {\bf Q} {\bf R}_i}
\end{eqnarray*}
${\bf m}_i$ and ${\bf R}_i$ denote the magnetic moment and position of the $i^{\rm th}$ spin in the unit cell. The method that was developed in this work involves the minimization of a $\chi^2$ defined as:
\begin{equation}
\label{chi2}
\chi^2({\bf m} ,H) = \sum_n \left(J^{\mbox{{\small calc}}}({\bf Q}_n, {\bf m} ) - I^{\mbox{{\small exp}}}({\bf Q}_n,H)\right)^2
\end{equation}
$\left\{ {\bf m} \right\}$ is the set of parameters to be refined. The $n$ index runs over a list of ${\bf Q}$ vectors given in Table \ref{table_Q} and which were chosen carefully as they show a significant evolution vs field. $ I^{\mbox{{\small exp}}}({\bf Q}_n,H)$ is the experimental intensity.  $J^{\mbox{{\small calc}}}({\bf Q}_n, {\bf m} )$ is derived from the calculated intensity $I^{\mbox{{\small calc}}}({\bf Q})$ but is defined using the moments ${\bf m} _1$ and ${\bf m} _2$ determined with {\sc Fullprof} at two reference fields $H_1 = 0$ and $\mu_0 H_2=\pm 1$ T: 
\begin{eqnarray*}
J^{\mbox{{\small calc}}}({\bf Q}_n, {\bf m} ) &=&
I^{\mbox{{\small exp}}}({\bf Q}_n, H_1)~+~
\alpha_n~r_n \\
r_n &=&
I^{\mbox{{\small calc}}}({\bf Q}_n,{\bf m} ) - I^{\mbox{{\small calc}}}({\bf Q}_n, {\bf m} _1) \\
\alpha_n &=& \frac{
I^{\mbox{{\small exp}}}({\bf Q}_n, H_2)-
I^{\mbox{{\small exp}}}({\bf Q}_n, H_1)
}
{
I^{\mbox{{\small calc}}}({\bf Q}_n, {\bf m} _2)-
I^{\mbox{{\small calc}}}({\bf Q}_n, {\bf m} _1)
}
\end{eqnarray*}
This definition ensures that 
\begin{equation*}
J^{\mbox{{\small calc}}}({\bf Q}_n, {\bf m} _1) = I^{\mbox{{\small exp}}}({\bf Q}_n, H_1)
\end{equation*}
and 
\begin{equation*}
J^{\mbox{{\small calc}}}({\bf Q}_n,{\bf m} _2) = I^{\mbox{{\small exp}}}({\bf Q}_n, H_2)
\end{equation*}
so that ${\bf m} _1$ and ${\bf m} _2$ minimize $\chi^2$ at $H_1$ and $H_2$ respectively. This definition thus puts severe constraints on the minimization. 

A home made program was then used to calculate $\chi^2({\bf m} ,H)$ over a grid for the parameters $\left\{ {\bf m} \right\}$, and to keep the values which minimize it. As different combinations of parameters can give a consistent result, a weight is associated to each solution and takes the form of a color scale in the figures presented in the main text: from blue, the worst $\chi^2$, to red, the best $\chi^2$. It is worth mentioning that this analysis cannot distinguish positive and negative values of the magnetic moments. A new criterion was thus added to keep only the values giving a positive product of the magnetization times the applied field (this is justified since no remanent magnetization has been observed in the magnetization curves). The points obtained with {\sc FullProf} from the data collections at fixed fields eventually confirm the consistency of the analysis. 

%%%%%%%%%%%%%%%%%%%%%%%%%%%%%%%%%%%%%%%%%%%%%%%%%%%%%%%%%%%%%%%%%%%

\section{Energy landscapes}
\label{landscape}

Following Ref. \cite{Xu_2019}, it is instructive to calculate the energy landscape in the presence of a magnetic field to better understand the origin of the hysteresis. Writing the four spins in a tetrahedron in their local frame $({\bf \tilde{x}},{\bf \tilde{z}})$ as:
\begin{equation}
\tilde{\tau} = (\sin \Phi_i , 0 , \cos \Phi_i)
\end{equation}
the classical exchange and Zeeman energy per tetrahedron can be evaluated using the Hamiltonian given by Eq. \ref{ham} and Eq. \ref{zeeman} (and neglecting the octupolar exchange ${\tilde {\sf J}}_{\tilde y}$ term). 
The numerical values $\tilde{{\sf J}}_{\tilde z} \approx -0.53$ K, $\tilde{{\sf J}}_{\tilde x} \approx$ 1.18 K, $\theta=1.23$ rad and  $\tilde{{\sf J}}_{\tilde z} \approx -0.53$ K, $\tilde{{\sf J}}_{\tilde x} \approx 0.97$ K, $\theta=1.08$ rad were used, corresponding to the pure and the 2.5 \% substituted sample parameters respectively.
% along with the numerical values $\tilde{{\sf J}}_{z} \approx$ -0.5 K and $\tilde{{\sf J}}_{x} \approx$ 1K.

\subsection{Field along $[001]$}

\begin{figure}[h!]
\includegraphics[width=8.2cm]{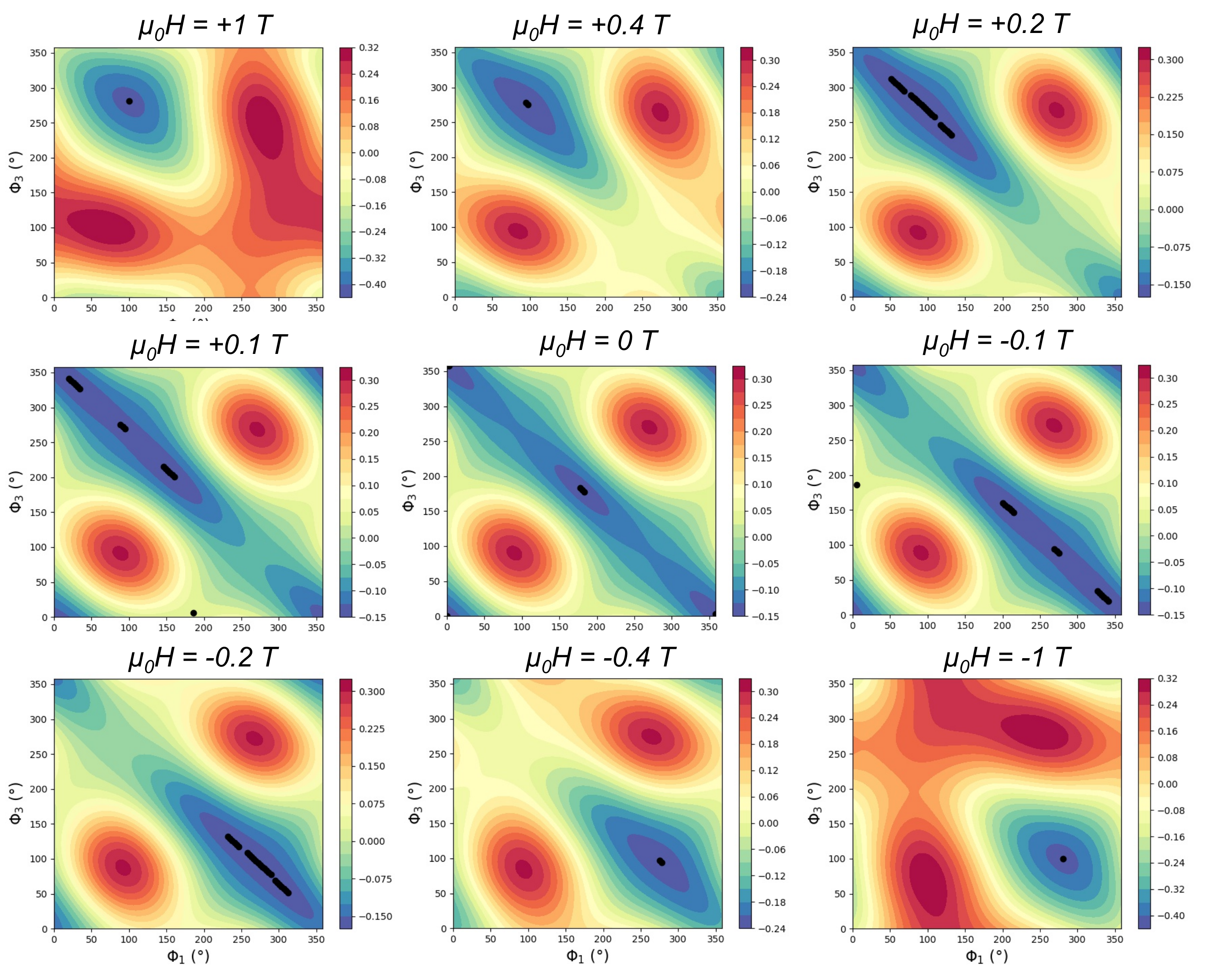}
\caption{\label{carte-energie} Contour plot of the classical energy ${\cal E}$ as a function of $\Phi_1$ and $\Phi_3$ for ${\bf H} \parallel [001]$ with the pure sample exchange parameters. The black dots show the positions of the minima and the color scale displays the energy in meV.}
\end{figure}

In the particular case of ${\bf H} \parallel [001]$, the angles $\Phi_i$ are identical within the two subgroups (1,2) and (3,4), hence: 
\begin{eqnarray*}
{\cal E} &=& g_z \mu_{\rm B} H \frac{2}{\sqrt{3}}\left( \cos(\Phi_1+\theta) - \cos (\Phi_3+\theta) \right) \\
&+& 2\tilde{{\sf J}}_{z} \left( \cos^2 \Phi_1 + 4\cos \Phi_1 \cos \Phi_3 + \cos^2 \Phi_3 \right) \\
&+& 2\tilde{{\sf J}}_{x} \left( \sin^2 \Phi_1 + 4\sin \Phi_1 \sin \Phi_3 +\sin^2 \Phi_3 \right)
\end{eqnarray*}
At large positive field, the minimum of the energy is obtained for $\Phi_1+\theta=\pi, \Phi_3+\theta=0$, i.e. $\Phi_1=\pi-\theta, \Phi_3=-\theta$ which corresponds to the moments of the (3,4) atoms parallel to their ${\bf z}$ axes and the (1,2) ones anti-parallel to their ${\bf z}$ axes. At zero field, two minima occur, corresponding to $\Phi_1=\Phi_3=0$ or $\pi$, i.e. to the domains ${\rm \overline{AIAO}}$ and ${\rm \overline{AOAI}}$ of the ${\bf \tilde z}$ ordered state. Figure \ref{carte-energie} illustrates this evolution. It displays several contour plots of ${\cal E}$ as a function of $\Phi_1$ and $\Phi_3$ for the ${\bf H} \parallel [001]$ case. The black dots show the position of the minima. Upon increasing field, two minima appear, separate to give rise to the ${\rm \overline{AIAO}}$ and ${\rm \overline{AOAI}}$ domains, and finally reconnect at a higher field. With a sufficiently large field, a single solution survives. 

\subsection{Field along $[111]$}

For ${\bf H} \parallel [111]$, the angles $\Phi_i$ are identical for the kagom\'e spins and we are left with two variables $\Phi_1$ (kagom\'e) and $\Phi_2$ (apical). The classical energy writes: 
\begin{eqnarray*}
{\cal E} &=& g_z \mu_{\rm B} H \left( -\cos(\Phi_1+\theta) + \cos(\Phi_2+\theta) \right) \\
&+& 6 \tilde{{\sf J}}_{z} \left( \cos^2 \Phi_1 + \cos \Phi_1 \cos \Phi_2 \right) \\
&+& 6 \tilde{{\sf J}}_{x} \left( \sin^2 \Phi_1 + \sin \Phi_1 \sin \Phi_2 \right)
\end{eqnarray*}
At large positive field, the minimum of the energy is obtained for $\Phi_1+\theta=0, \Phi_2+\theta=\pi$, i.e. $\Phi_1=-\theta, \Phi_2=\pi-\theta$ which corresponds to spins 1, 3, 4 parallel to their ${\bf z}$ axes and spin 2 anti-parallel to its ${\bf z}$ axis. At zero field, two minima occur, corresponding to $\Phi_1=\Phi_3=0$ or $\pi$, i.e. to the two domains of the AIAO configuration.

\subsection{Field along $[1\bar{1}0]$}

For ${\bf H} \parallel [1{\bar 1}0]$, two different regimes can be considered. At large fields, $|H| \geq H_{110}$, $\Phi_2=\pi+\Phi_1$ and $\Phi_3=\pi+\Phi_4$, hence the classical energy:
\begin{eqnarray*}
{\cal E} &=& g_z \mu_{\rm B} H 2\sqrt{\frac{2}{3}} \cos(\Phi_3+\theta) \\
&-& 2\tilde{{\sf J}}_{z} \left( \cos^2 \Phi_1 + \cos^2 \Phi_3 \right) \\
&-& 2\tilde{{\sf J}}_{x} \left( \sin^2 \Phi_1 + \sin^3 \Phi_3 \right)
\end{eqnarray*}
At large positive field, the minimum of the energy is obtained for $\Phi_3+\theta=\pi$, i.e. $\Phi_3=\pi-\theta, \Phi_4=-\theta$ which corresponds to spin 3 antiparallel to its ${\bf z}$ axis and spin 4 parallel to its ${\bf z}$ axis. 

At small fields, $|H| \leq H_{110}$, no simplification appears but $\Phi_1=\Phi_2$ and we are left with three independent variables $\Phi_{1,3,4}$. The classical energy writes: 
\begin{widetext}
\begin{eqnarray*}
{\cal E} &=& g_z \mu_{\rm B} H \sqrt{\frac{2}{3}} \left( \cos(\Phi_3+\theta) - \cos(\Phi_4+\theta) \right) \\
&+& 2\tilde{{\sf J}}_{z} \left( \cos^2 \Phi_1 + 2\cos \Phi_1 (\cos \Phi_3+\cos \Phi_4) +\cos \Phi_3 \cos \Phi_4 \right) \\
&+& 2\tilde{{\sf J}}_{x} \left( \sin^2 \Phi_1 + 2\sin \Phi_1 (\sin \Phi_3+\sin \Phi_4) +\sin \Phi_3 \sin \Phi_4 \right)
\end{eqnarray*}
\end{widetext}
The minimization of ${\cal E}$ in the general case should be carried out in the three-dimensional space spanned by the three $\Phi_i$ angles. 

%The mean field phase diagrams shown in Figure \ref{MF_001-110}, \ref{MF_111_FC} and \ref{MF_111_hyst} are calculated using a mean field approximation of Eq. \ref{ham} and Eq. \ref{zeeman}. 

%%%%%%%%%%%%%%%%%%%%%%%%%%%%%

%%%%%%%%%%%%%%%%%%%%%%%%%%%%%%%%%%%%%%%%
%Biblio
%%%%%%%%%%%%%%%%%%%%%%%%%%%%%%%%%%%%%%%%
\bibliography{biblio}

\end{document}